\documentclass[referee]{raa}            

\usepackage{graphicx,times}             
\usepackage{natbib}
\usepackage{amssymb,amsmath}
\bibpunct{(}{)}{;}{a}{}{,}
\usepackage{indentfirst}
\usepackage[pagebackref=true]{hyperref}
\usepackage{caption}
\usepackage[utf8]{inputenc}
\DeclareUnicodeCharacter{02BC}{'}
\DeclareUnicodeCharacter{2212}{-}
\usepackage{rotating} 
\usepackage{longtable}

\begin{document}

  \title{Polarization Profiles of Globular Cluster Pulsars from FAST I. 25 Profiles from Previously Known Pulsars}

   \volnopage{Vol.0 (20xx) No.0, 000--000}      
   \setcounter{page}{1}          

   \author{Tong Liu 
      \inst{1,2}
   \and Lin Wang
      \inst{3}\thanks{*Corresponding author, email: linwang@pku.edu.cn, panzc@bao.ac.cn}
   \and Zhichen Pan
      \inst{1,2,4,5}
   }
   \institute{National Astronomical Observatories, Chinese Academy of Sciences, 20A Datun Road, Chaoyang District, Beijing 100101, Peopleʼs Republic of China \\
        \and
             College of Astronomy and Space Sciences, University of Chinese Academy of Sciences, Beijing 100049, People's Republic of China\\
        \and
             Shanghai Astronomical Observatory, Chinese Academy of Sciences, 80 Nandan Road, Shanghai 200030, People's Republic of China\\
        \and
             CAS Key Laboratory of FAST, National Astronomical Observatories, Chinese Academy of Sciences, Beijing 100101, People's Republic of China; {\it panzc@bao.ac.cn}\\
        \and
             Guizhou Radio Astronomical Observatory, Guizhou University, Guiyang 550025, People's Republic of China\\
\vs\no
   {\small Received 20xx month day; accepted 20xx month day}}

\abstract{
Pulsar polarization profiles are critical for understanding their magnetospheric structures and radiation mechanisms. 
We present polarization profile measurements for 25 pulsars in globular clusters (GCs) from the observation of the Five-hundred-meter Aperture Spherical radio Telescope (FAST).
The diversity of polarization profiles shows complex magnetic structure and emission pattern of pulsars.
Among these, the polarization profiles of 15 pulsars were firstly measured.
M53A present a 57\% linear polarization ratio, being highest among these 25 GC pulsars.
M15H present a 42\% circular polarization ratio, which is the highest among these 25 pulsars.
The average ratios of circular and absolute circular polarization of these GC pulsars are -1\% and 10\% respectively, 
lower than normal pulsars measured with Parkes (\citealt{Oswald2023}), which are 5\% and 32\%, respectively.
The Rotation Measure (RM) values of each pulsar were measured, 
giving the range of pulsar(s) from M3, M5, M15, M71, M53, NGC6517, NGC6539, NGC6760 being
8(1) to 16(1) rad $m^{-2}$, 
-1(2) to -4(6) rad $m^{-2}$, 
-70(2) to -76(1) rad $m^{-2}$,
-480(14) rad $m^{-2}$ (M71A only),
-2(1) rad $m^{-2}$ (M53A only),
187(1) to 212(2) rad $m^{-2}$, 
109(1) rad $m^{-2}$ (NGC6539A only), 
and 102(1) to 129(4) rad $m^{-2}$.
The GCs closer to the Galactic plane (GP) tend to have larger RM.
This is consistent with previous study (\citealt{Hutschenreuter2022}).}


    \authorrunning{Tong Liu, Lin Wang \& Zhichen Pan }            
    \titlerunning{Polarization Profiles of GC Pulsars from FAST}  

\maketitle
\section{Introduction}
\label{sect:intro}

Pulsars are among the most highly polarized astrophysical sources in the universe \citep{Lyne1968}. These polarization properties and the shape of their profiles are closely tied to the radiation mechanisms and magnetic field structures of pulsars.

For instance, the rotation vector model (RVM; \citep{Radhakrishnan1969}) offers a theoretical framework for interpreting the variation of polarization position angles (PAs) across pulse phases, particularly in slow pulsars. These pulsars, typically characterized by spin periods exceeding 30 ms and often undergoing a recycling phase. This model is grounded in the geometry of the pulsar's magnetic field and its orientation relative to the emission beam, providing critical insights into the emission mechanisms and magnetic field structure of these objects.

Expanding upon this, the core and cone model \citep{Lyne1988} explains the observed polarization and intensity profiles in terms of core and conal emission components within the pulsar's radio beam. Furthermore, the patchy beam model \citep{Karastergiou2007} explores the relationship between the radial emission height and the resulting pulse shape, offering insights into the complex structure of pulsar emission regions. 

The development of these models has primarily been based on observations of slow pulsars and millisecond pulsars (MSPs) located in the Galactic plane (GP). However, pulsars residing in globular clusters (GCs) often experience unique evolutionary processes due to the highly dynamic and dense environments within GCs, including frequent stellar collisions and gravitational interactions \citep{Rappaport1989}. This raises an interesting question: do pulsars in GCs exhibit polarization properties similar to those outside GCs? Current understanding remains insufficient to conclusively address this question, highlighting the need for further investigation into the polarization characteristics of GC pulsars.

Rotation measure (RM) measurements of pulsars across a wide range of Galactic locations provide a valuable tool for mapping the large-scale structure of the Galactic magnetic field \citep{Han2006}. Unlike pulsars located in the GP, GCs are typically situated in the Galactic halo. Thus, measuring the RMs of pulsars within GCs offers a unique opportunity to probe the magnetic field distribution in the halo, thereby enhancing our understanding of the Galactic magnetic field's three-dimensional structure \citep{Abbate2020}.

%
%
%
Since the discovery of the first GC pulsar, J1824-2452A, by \citet{Lyne1987}, over 340 pulsars have been reported in 45 GCs\footnote{\url{https://www3.mpifr-bonn.mpg.de/staff/pfreire/GCpsr.html}}. These pulsars provide opportunities to the study of GC evolution and interstellar medium (e.g., \citealt{Benacquista2013}). However, GC pulsars are generally more distant and fainter than those in the GP, making the measurement of their polarization properties challenging. High-sensitivity observations are therefore essential to accurately characterize the polarization features of these pulsars.

%

The advent of the Five-hundred-meter Aperture Spherical radio Telescope (FAST; \citealt{NAN2011}) has significantly enhanced the sensitivity of GC pulsar observations.
Among the 93 pulsars discovered in 17 GCs within the FAST sky, 
more than two-thirds were discovered by FAST\footnote{\url{e.g., see, https://fast.bao.ac.cn/cms/article/65/}}.
Timing solutions have been made for more than 70 pulsars such as the recently reported in NGC6517 (\citealt{yindejiang_6517}) and M3 (\citealt{libaoda_m3}).
However, only 11 polarization profiles of these 93 pulsars have been measured to date (e.g., \citealt{Wang2023} and
\citealt{Pan2023}).

Here we present polarization profile measurements of 25 pulsars in GCs, including 15 measured for the first time as our first results from continuing timing and monitoring GC pulsars with FAST.
All these 25 pulsars were discovered before FAST.
Our work provides samples for studying pulsar magnetic field structures, emission mechanisms, and the magnetic field properties of GCs. 
Observations and data acquisition are described in Section 2, with data reduction in Section 3. 
Results and discussion are presented in Section 4, followed by conclusions given in Section 5.


\section{Observations and Data Acquisition}
\label{sect:Obs}

In FAST sky, 27 GC pulsars were discovered before FAST construction completion. 
Most of these pulsars have public timing solutions and can be relatively bright for FAST to obtain their polarization profiles.
As our first phase of work, we processed the data of 25 pulsars from 8 GCs including M53 (M53A), M3 (M3A, B, and D), M5 (M5A to E), NGC6517 (NGC6517A to D), NGC6539 (NGC6739A), NGC6760 (NGC6760A and B), M71 (M71A) and M15 (M15A to H).
The two left are M3C and NGC 6749B, which were never confirmed.

The data used in this work were observed using the 19-beam receiver of FAST, 
covering a frequency range of 1 to 1.5 GHz . 
The signal were channelized and digitized using a Reconfigurable Open Architecture Computing Hardware (ROACH)\footnote{developed by the Collaboration for Astronomy Signal Processing and Electronics Research (CASPER) group; http://casper.berkeley.edu/} unit. 
Then, the data were subsequently packetized and stored in search mode PSRFITS format including four polarizations (AA, BB, AB, BA). 
The channel bandwidth and sampling time of the data were 122\,kHz and a 49.152 $\mu s$, respectively. 
The detailed information of FAST receiver system, such as system temperature, sky coverage, etc., 
can be found in \citet{Jiang2020}.
At the beginning of each observation, a 1-min noise diode signal were injected with period of 0.2\,s. 
The integration of each observation were decided by the Declination of the GCs in the sky.
We observed each GC as longer time as we can to achieve higher signal-to-noise ratios (S/Ns). 
The source and observation details can be found in see Tab \ref{tab1}.

\begin{table}
\begin{center}
\caption[]{FAST GC Observations and Data References for Polarization Measurements}\label{tab1}
\begin{tabular}{lcccccc}
    \hline\hline\noalign{\smallskip}
    GC & Date & RA & Dec & Duration & PSR & Reference \\
    \hline\noalign{\smallskip}
    M53 & 20241007 & 13:12:55.00 & +18:10:05.4 & 12900 & M53A & \citealt{lianyujie_m71} \\
    \hline\noalign{\smallskip}
    M3  & 20241003 & 13:42:12.00 & +28:22:38.2 & 9900  & M3B & \citealt{libaoda_m3} \\
        & 20240921 & 13:42:12.00 & +28:22:38.2 & 10995 & M3A & \\
        & 20240924 & 13:42:12.00 & +28:22:38.2 & 11775 & M3D & \\
    \hline\noalign{\smallskip}
    M5  & 20220216 & 15:18:33.22 & +02:04:51.7 & 7200  & M5C, D & \citealt{zhanglei_m5} \\
        & 20220211 & 15:18:33.22 & +02:04:51.7 & 7200  & M5A, B, E & \\
        
    \hline\noalign{\smallskip}
    NGC6517 & 20230929 & 18:01:51.52 & -08:57:31.6 & 9000 & NGC6517D & \citealt{yindejiang_6517} \\
            & 20221231 & 18:01:51.00 & -08:57:31.6 & 8400 & NGC6517A, B& \\
            & 20221228 & 18:01:51.00 & -08:57:31.6 & 8400 & NGC6517C & \\
    \hline\noalign{\smallskip}
    NGC6539 & 20241009 & 18:04:50.00 & -07:35:09.1 & 3000 & NGC6539A & \citealt{Hobbs2004} \\
    \hline\noalign{\smallskip}
    NGC6760 & 20240930 & 19:11:02.1  & +01:01:49.7 & 6600 & NGC6760A, B & \citealt{Freire2005} \\
    \hline\noalign{\smallskip}
    M71     & 20220530 & 19:53:44.00 & +18:44:31.0 & 5285 & M71A & Lian et al. in prep \\
    \hline\noalign{\smallskip}
    M15     & 20231206 & 21:29:58.00 & +12:10:01.2 & 13560 & M15A to D & \citealt{wuyuxiao_m15} \\
            & 20220428 & 21:29:58.25 & +12:10:01.23 & 7200 & M15E to H & \\
    \hline\hline\noalign{\smallskip}
\end{tabular}
\end{center}
\end{table}


\section{Data Reduction}
\label{sect:data}

\subsection{Polarization Calibration}

The signal from both noise diode and pulsars were folded into sub-integrations using the software package \textsc{DSPSR} \citep{vanStraten2011}.
The length of the sub-intergration is 60~s.
The number of phase bin of search pulse is 128.
For most of these 25 pulsars, the ephemerides we obtained from either {\tt psrcat}  \citep{Manchester2005} \footnote{\url{https://www.atnf.csiro.au/research/pulsar/psrcat/}} or publications are accurate enough for our FAST data.
However, NGC6760A presented a spinning period shifting.
So, the routine {\tt pdmp} from \textsc{PSRCHIVE} was used to search for the spinning period.
Since the effective bandwidth of 19-beam receiver is approximately 1.05 to 1.45\,GHz, 
we removed the bands of 1.00 to 1.05\,GHz and 1.450 to 1.50\,GHz using the {\tt pazi} routine of \textsc{PSRCHIVE} \citep{Hotan2004}. 
According to the previously known radio frequency interferences (RFIs),
the data in the frequency range of 1065 to 1085, 1085 to 1095, 1200 to 1215, and 1260–1285\,MHz were also removed. 
Besides, we also manually removed the obvious RFIs in time and frequency domains.
In order to improve S/N in each frequency band, we scrunched the folded data into 256 subbands.

%
%
%
%
%

%
%
%
%
We subsequently conducted a standard polarization calibration procedure to account for the contributions of the telescope and observing system to the measured polarization. These effects are typically characterized by the Mueller matrix, M, as described in \citet{Lorimer2005} and represented in Equation \ref{eq:muelller}. 

\begin{equation}\label{eq:muelller}
\begin{aligned}
M &= M_{amp} \times M_{cc}  \times M_{PA}.
\end{aligned}
\end{equation}

The Mueller matrix components were used to characterize specific instrumental and observational effects. The component M$_{PA}$ accounts for the effect of the parallactic angle, with the feed of FAST fixed relative to the plane of polarization. M$_{cc}$ represents the cross-coupling effect arising from the two orthogonal probes of the receiver. Given the minimal leakage between the two feeds of the 19-beam receiver, the single-axis model is accurate to within 0.04\% \citep{ching2022}. Finally, 
M$_{amp}$ accounts for variations in gain and phase introduced by differences in the amplifier chains through which the signals pass.
The polarisation calibration was performed via the {\tt pac} routine in \textsc{PSRCHIVE}.

RM values were calculated for all observations using the {\tt rmfit} routine in the \textsc{PSRCHIVE} software package. This method involves iteratively determining the differential position angle ($\Delta$PA) of the linear polarization by splitting the data bandwidth into two half-bands. The relationship between $\Delta$PA and RM is described by the equation:
\begin{equation}
    \Delta \rm PA = \lambda ^2 \times \rm RM .
	\label{eq:rm}
\end{equation}

%
%



\subsection{Polarization Parameter Acquisition}
We used {\tt pam} routine to sum the data both in time and frequency to obtain the mean pulse profiles with best S/N.
We then measured pulse widths at 10\% $W_{10}$  and 50\% $W_{50}$ of the peak intensity, respectively. To ensure accurate measurements, the pulse profiles were first smoothed using a Gaussian function. The on-pulse region was defined as the range of intensities exceeding three times the standard deviation of the baseline noise (3$\sigma$).

%
%

From each averaged profile, we derived all Stokes parameters (I, Q, U, V). The linear polarisation (L$_{i}$) for each phase bin was calculated as: 
$$ L_i =
\begin{cases}
\sqrt{L_{meas,i}^2 - (\sigma_{Q}^2 + \sigma_{U}^2)} & \text{if } L_{meas,i}^2 \geq \sigma_{Q}^2 + \sigma_{U}^2 \\
-\sqrt{(\sigma_{Q}^2 + \sigma_{U}^2) - L_{meas,i}^2} & \text{otherwise.}
\end{cases}
$$

%
%
%
%
Where, L$_i$ represents the linear polarization in phase bin 
after applying the Weisberg correction. The measured linear polarization, L$_{\rm meas}$ is defined as L$_{meas}^2$ = Q$_{meas}^2$+U$_{meas}^2$, where $\sigma_Q$ and $\sigma_U$ are the measured Stokes parameters. The uncertainties $\sigma_Q$ and $\sigma_U$ correspond to the root-mean-square (RMS) values of Q and U, respectively. PAs were calculated using the relationship ${\rm PA}=\frac{1}{2} \arctan(U/Q)$. The uncertainties in the PAs($\sigma_{PA}$) were determined through standard error propagation techniques. Only PAs with a confidence level greater than 3-$\sigma$ were retained.



\section{Results and Discussion}


\subsection{Pulse Width}

\begin{figure}[htbp!]
    \centering
    \begin{minipage}{0.45\textwidth}
        \includegraphics[width=\linewidth]{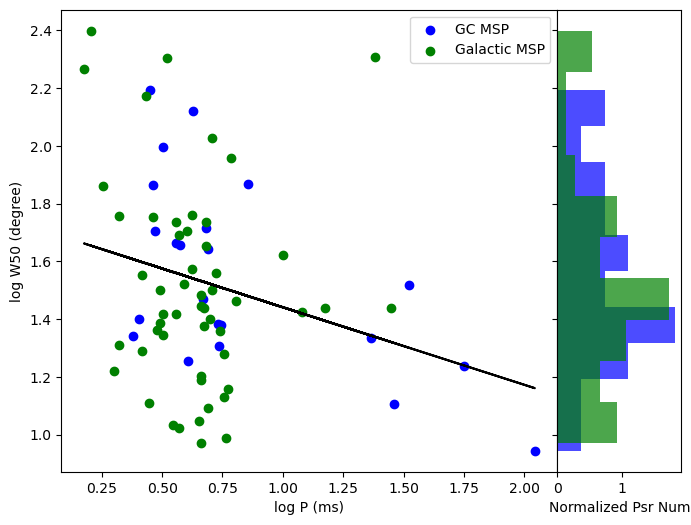}
    \end{minipage}%
    \hfill
    \begin{minipage}{0.45\textwidth}
        \includegraphics[width=\linewidth]{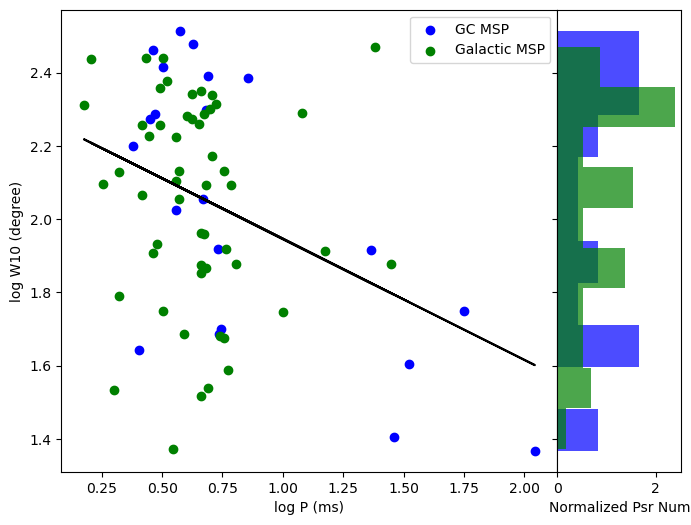}
    \end{minipage}%
    \vspace{0pt}
    \caption{Panels on left and right are relationship between pulse width $W_{50}$ or $W_{10}$ with the period for pulsars.
    Blue points and bars represent GC MSPs, and green points and bars represent Galactic MSPs from \citealt{Wang2023}.
    Solid lines represent power-law fits to the data.
    For comparison, histograms on the right of each panel are accumulated normalized pulsar numbers.}
\label{fig1}
\end{figure}

We compared the pulse width ($W_{10}$ and $W_{50}$) distributions of the 25 GC pulsars in this work with those of a sample of MSPs in the GP from \citet{Wang2023}. 
The comparison, shown in Figure~\ref{fig1}, reveals that the two pulsar populations exhibit similar distributions for both $W_{10}$ and $W_{50}$. 
Additionally, we examined the relationship between $W_{10}$ and $W_{50}$ with pulsar spin periods for GC pulsars and the same set of MSPs in the GP, 
as depicted in Figure~\ref{fig1}. 
A power-law fit was performed for both GC and GP MSPs, yielding power-law indices of -0.268 ($W_{10}$) and -0.330 ($W_{50}$), respectively.
For comparison, \citet{Karastergiou2024} reported a power-law index of -0.308 for the relationship between $W_{10}$ and spin period, 
based on a sample that included both MSPs and slow pulsars observed with MeerKAT. 
Our result are consistent with the result of \citet{Karastergiou2024}.


%

%


\subsection{Profiles and Polarization}

\label{sect:discussion}
The polarization profiles for the 25 GC pulsars were presented in Figure~\ref{fig2}. 
The corresponding polarization parameters were summarized in Table~\ref{tab2}, 
including the fractions of linear polarization, circular polarization, 
the absolute value of circular polarization, $W_{10}$ and $W_{50}$, and RM values.
Among them, 10 were measured by \citealt{Wang2023}.
Our results are consistent with their results. 
Notably, M15C is 
the only one residing in a double neutron star system. 
This pulsar exhibits a relatively long spin period (30.5\,ms) and is characterized by a single, narrow pulse in its mean pulse profile.
%
%
The polarization profile of M15C obtained from the 7200 second observation done in April 28, 2022, exhibits a lower S/N of 4,
in contrast to the significantly higher S/N reported in the 2019 profile (from a 7320-second observation) by \citet{wuyuxiao_m15}. 
This disparity in S/N suggests that M15C may experience significant variability.
This meets the prediction that the luminosity of M15C kept decreasing due to orbit variation (\citealt{Ridolfi2017}).

M5A, NGC6539A and M15E are the only three exhibiting orthogonal jump with PA.
%
M5A exhibits two linear polarization pulses within a single total intensity pulse.
It is likely that the total intensity pulse of M5A is composed of multiple overlapping peaks.
M5E exhibits a complex pulse profile, with linear polarization ratio in the third component reaching nearly 100\%.
Despite a 10.6 S/N of total intensity, M15H exhibits 37\% circular polarization ratio, highest of all pulsars in this work.
%
%
M53A and NGC6760A also exhibit high circular polarization ratio, being -21\% and 19\%, respectively.
Among the 25 GC pulsars, M5C, M5D, and M5E are the only three pulsars exhibiting inter-pulse emission, 
which is commonly attributed to radiation originating from the pulsar's opposite magnetic pole.
Considering that M5C is an eclipsing black widow system, 
we measured the polarization profiles for each 5-minute integration across different orbital phases. 
However, we did not observe any significant variations in the polarization characteristics at these phases, 
in contrast to the behavior of PSR J1720-0533 reported by \citet{Wang2021} for the black widow J1720−0533. 


\begin{figure}[h!]
    \centering
    \begin{minipage}{0.33\textwidth}
        \includegraphics[width=\linewidth]{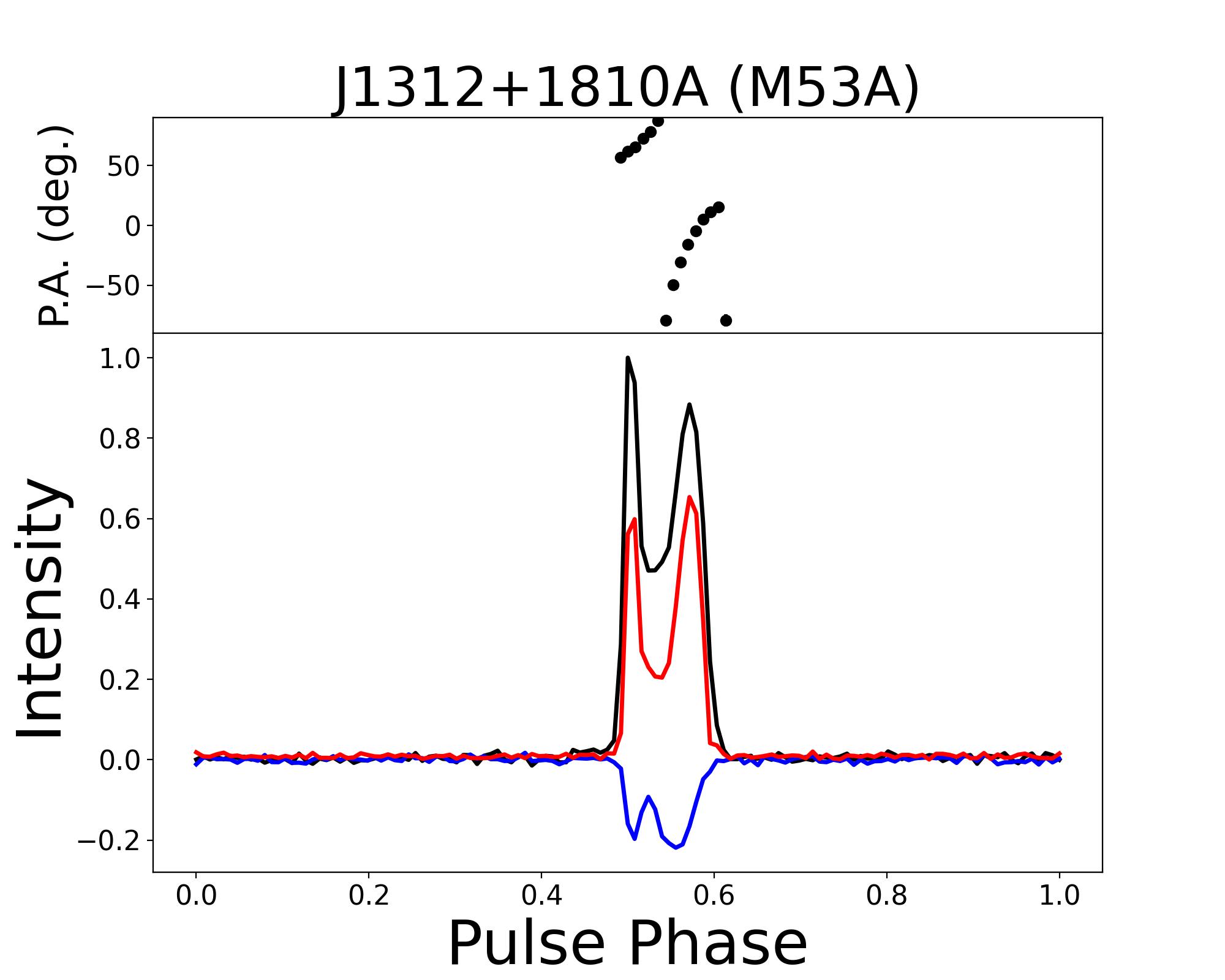}
    \end{minipage}%
    \hfill
    \begin{minipage}{0.33\textwidth}
        \includegraphics[width=\linewidth]{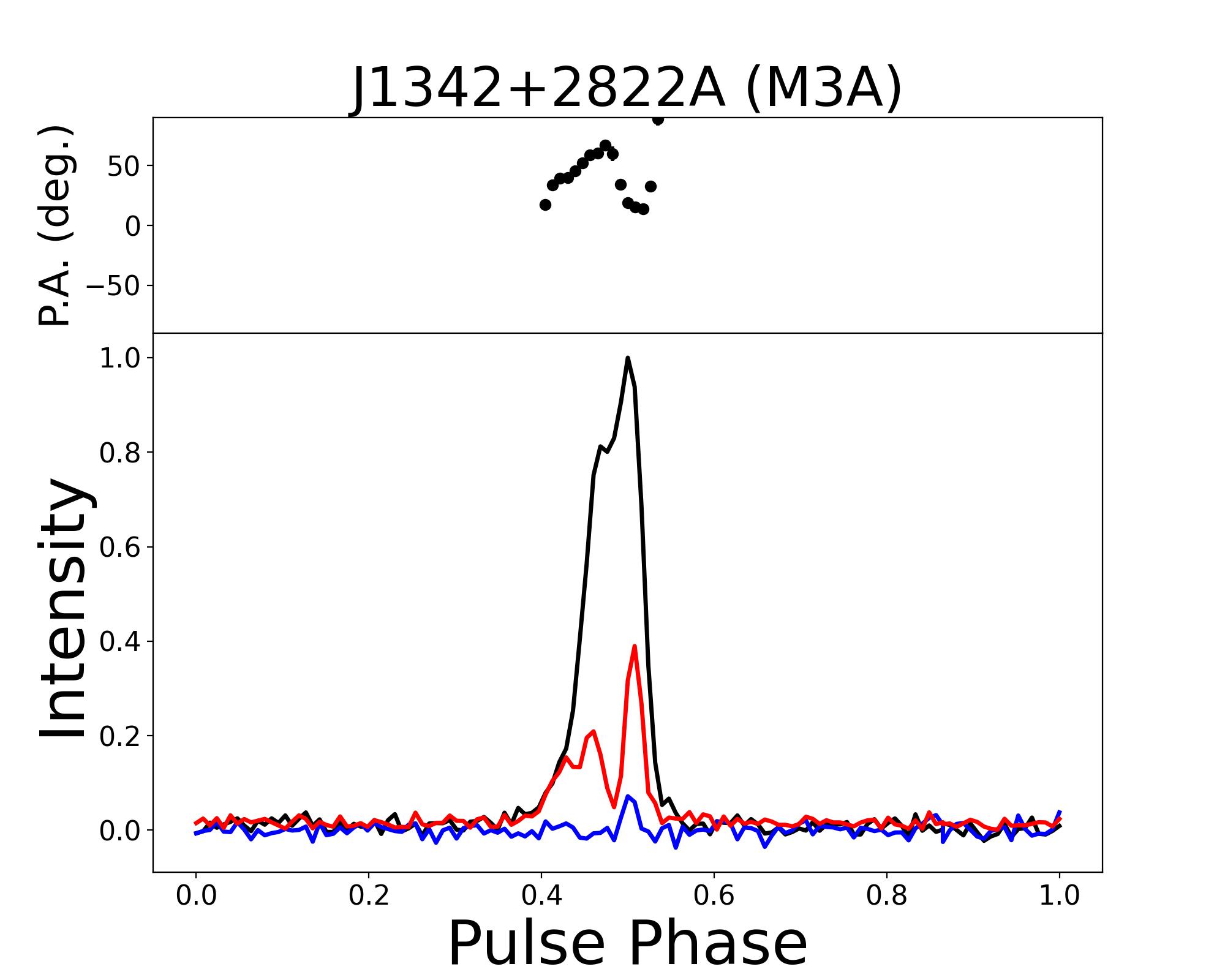}
    \end{minipage}%
    \hfill
    \begin{minipage}{0.33\textwidth}
        \includegraphics[width=\linewidth]{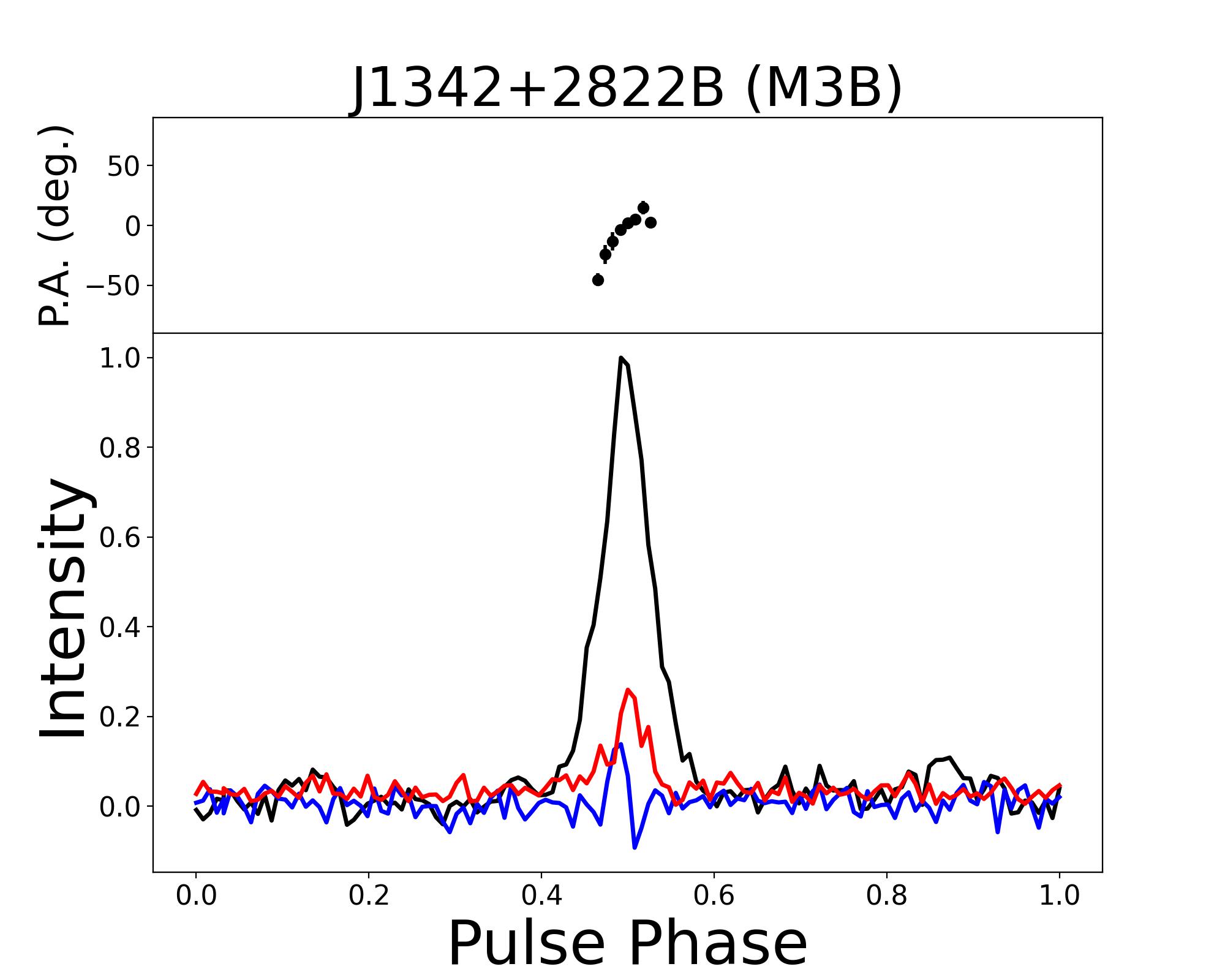}
    \end{minipage}%
    \vspace{5pt}
    \begin{minipage}{0.33\textwidth}
        \includegraphics[width=\linewidth]{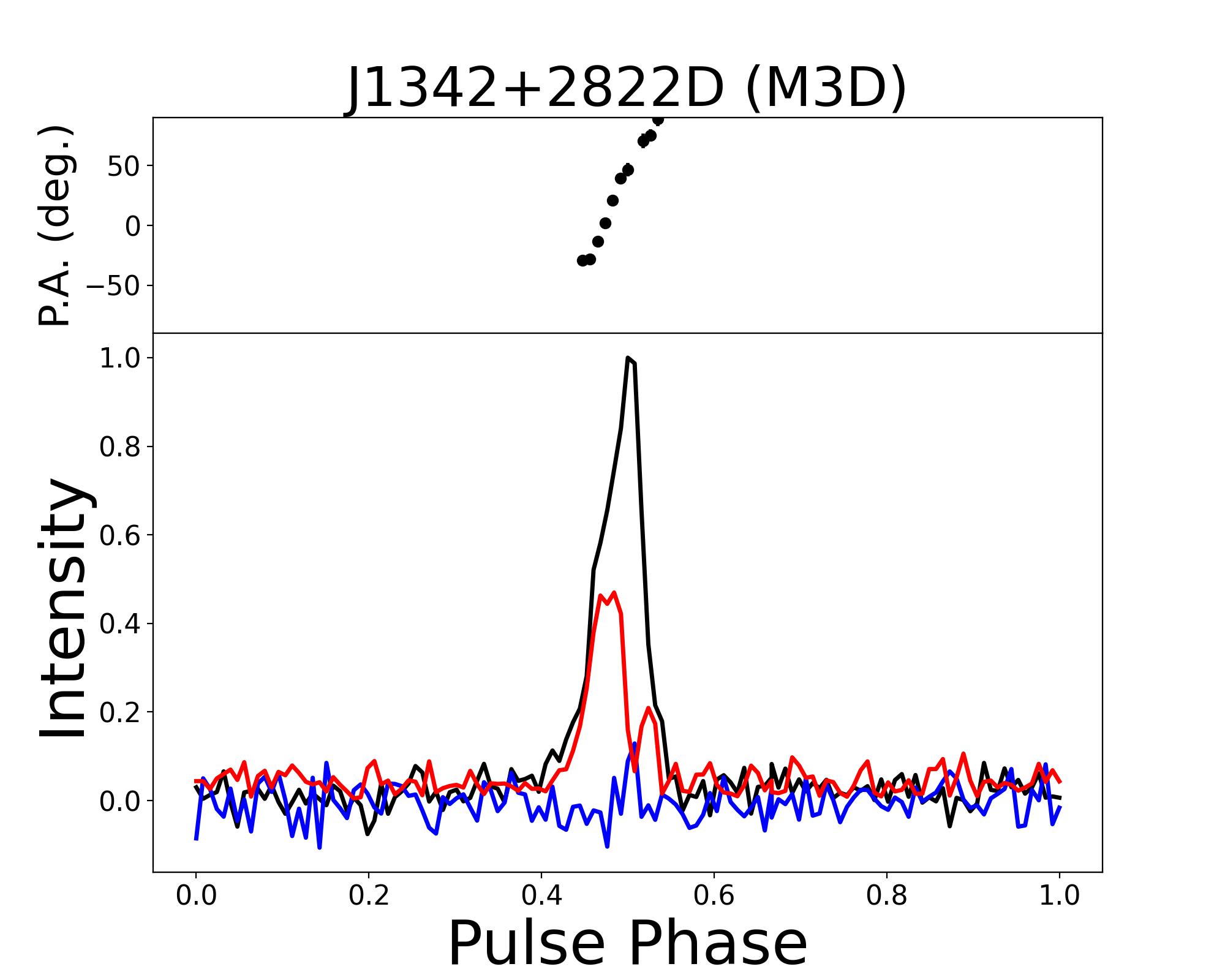}
    \end{minipage}%
    \hfill
    \begin{minipage}{0.33\textwidth}
        \includegraphics[width=\linewidth]{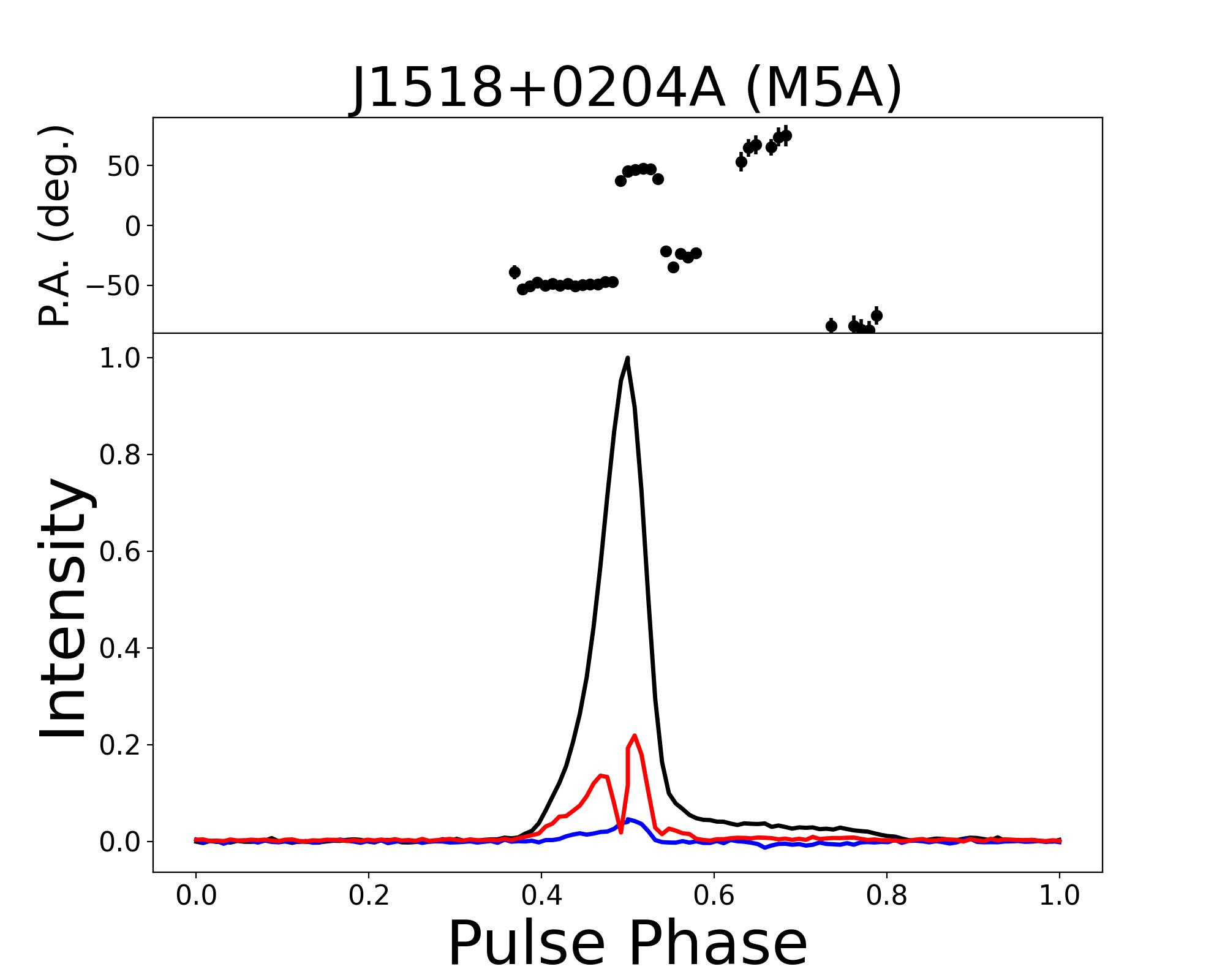}
    \end{minipage}%
    \hfill
    \begin{minipage}{0.33\textwidth}
        \includegraphics[width=\linewidth]{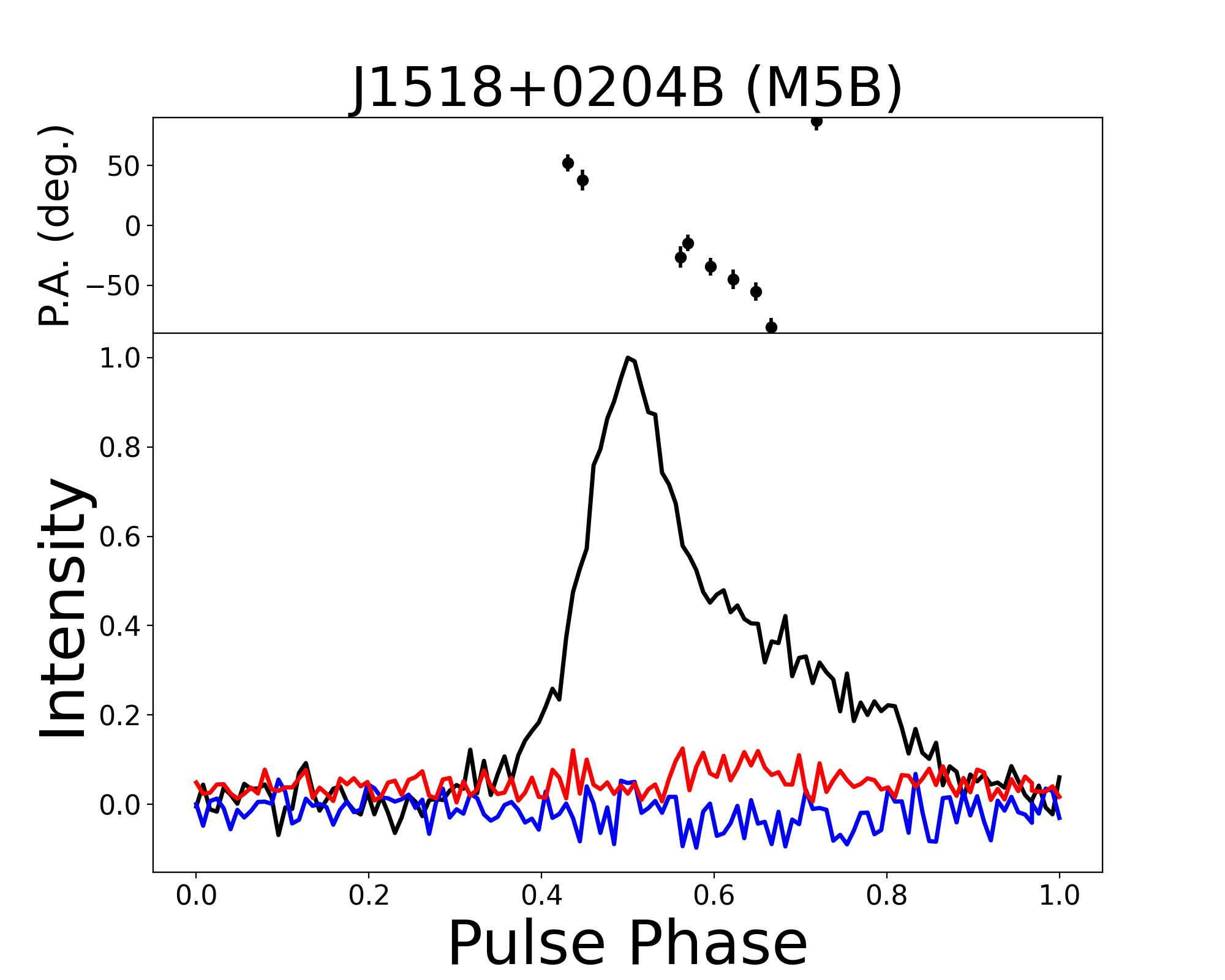}
    \end{minipage}%
    \vspace{5pt}
    \begin{minipage}{0.33\textwidth}
        \includegraphics[width=\linewidth]{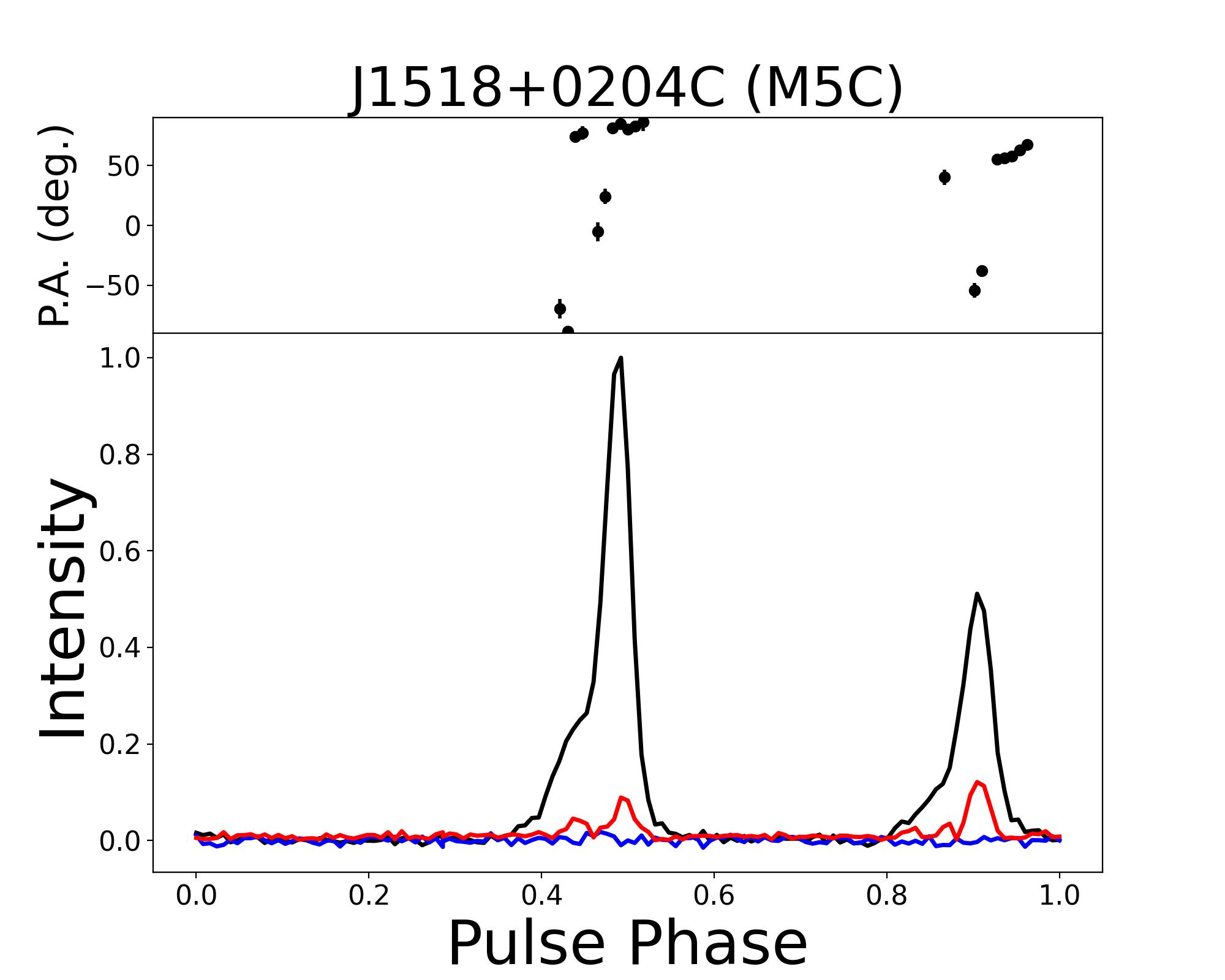}
    \end{minipage}%
    \hfill
    \begin{minipage}{0.33\textwidth}
        \includegraphics[width=\linewidth]{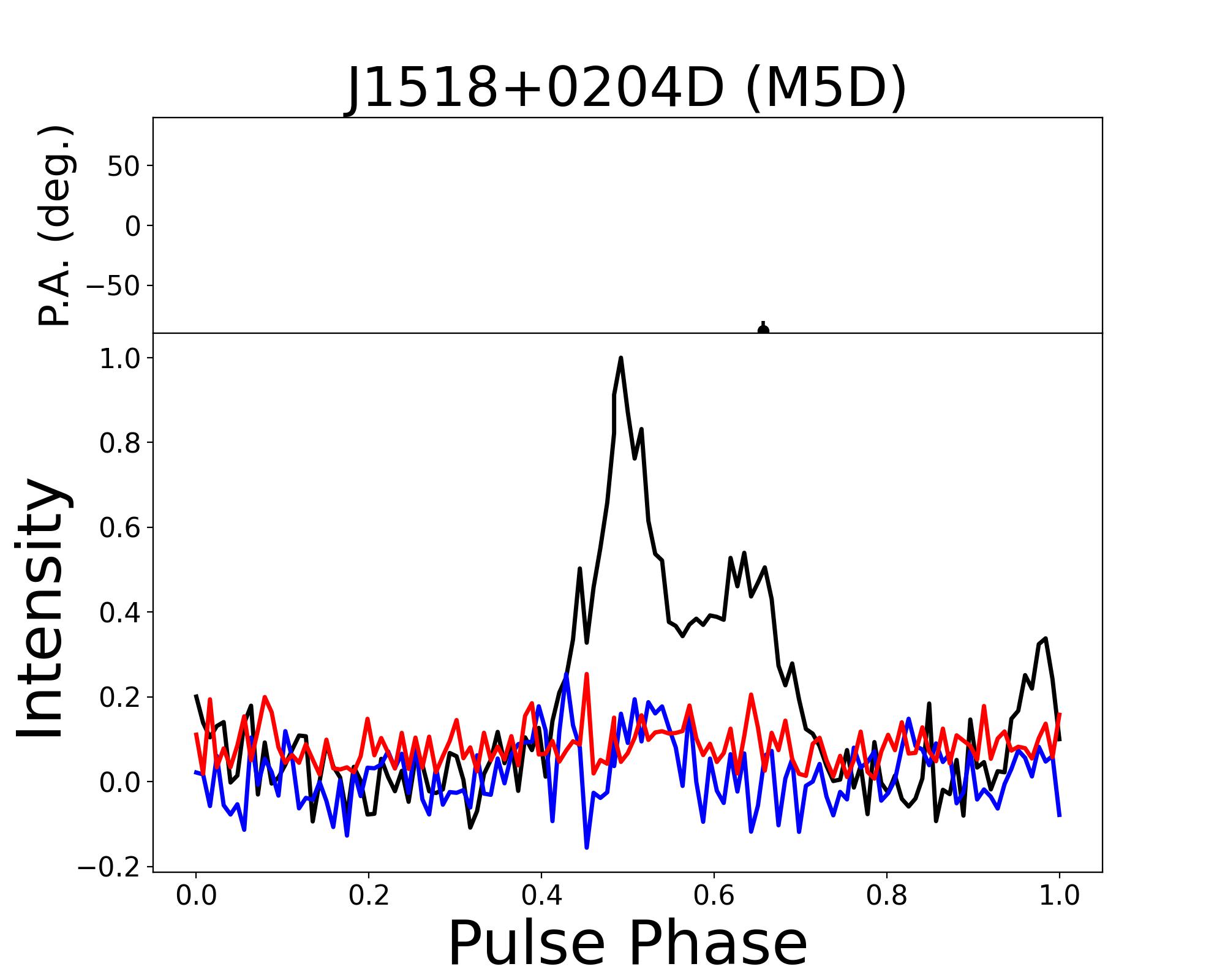}
    \end{minipage}%
    \hfill
    \begin{minipage}{0.33\textwidth}
        \includegraphics[width=\linewidth]{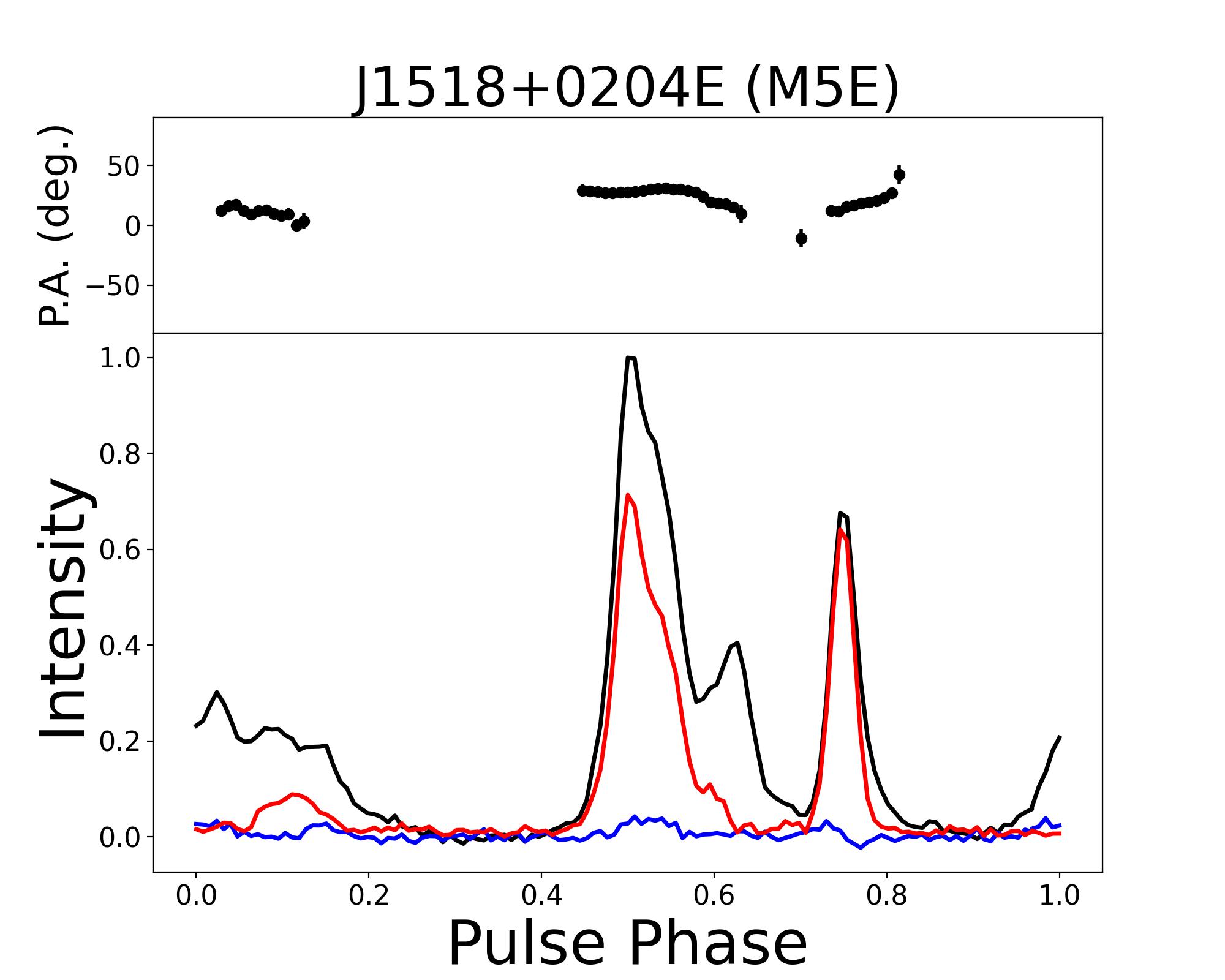}
    \end{minipage}%
    \vspace{5pt}
    \begin{minipage}{0.33\textwidth}
        \includegraphics[width=\linewidth]{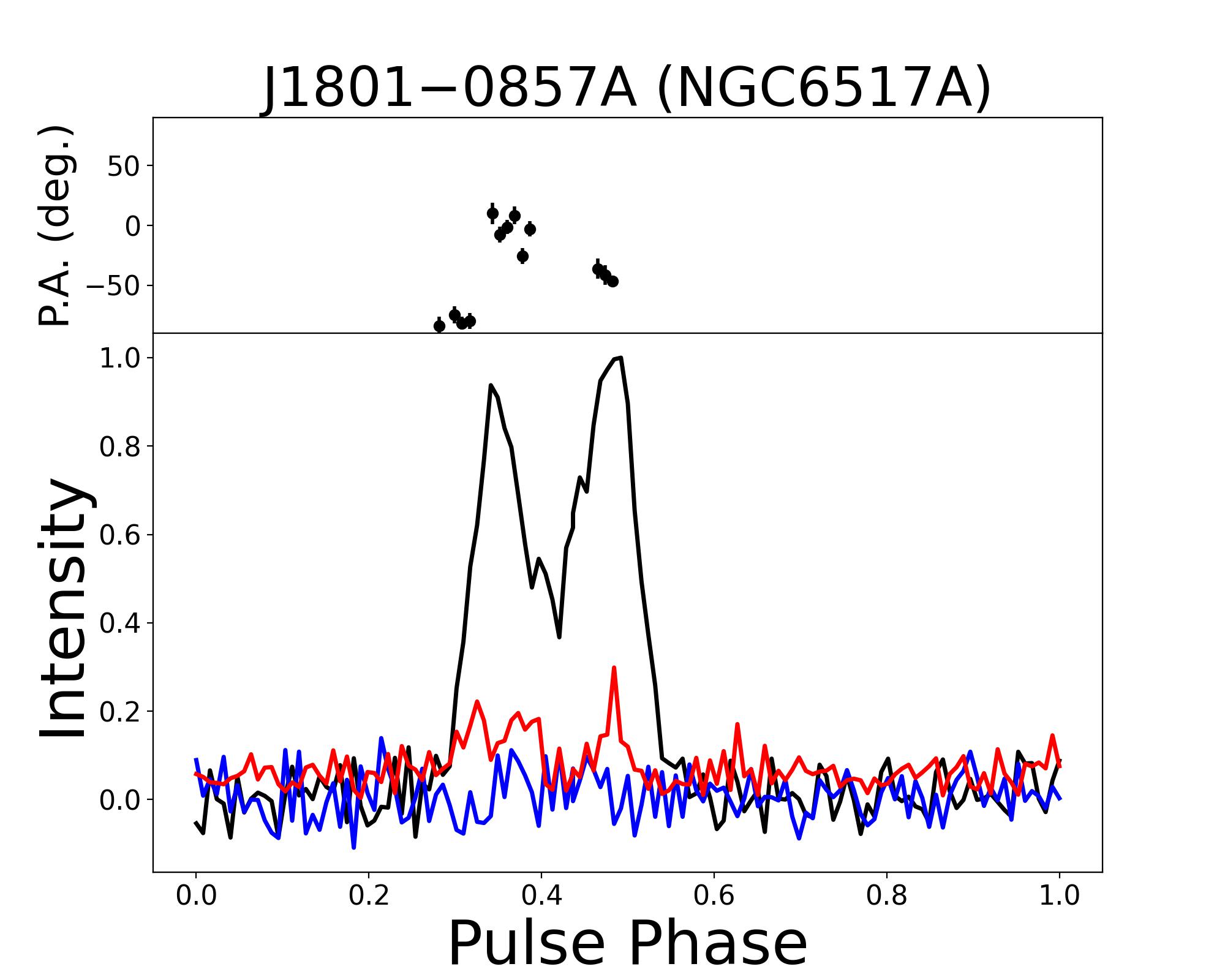}
    \end{minipage}%
    \hfill
    \begin{minipage}{0.33\textwidth}
        \includegraphics[width=\linewidth]{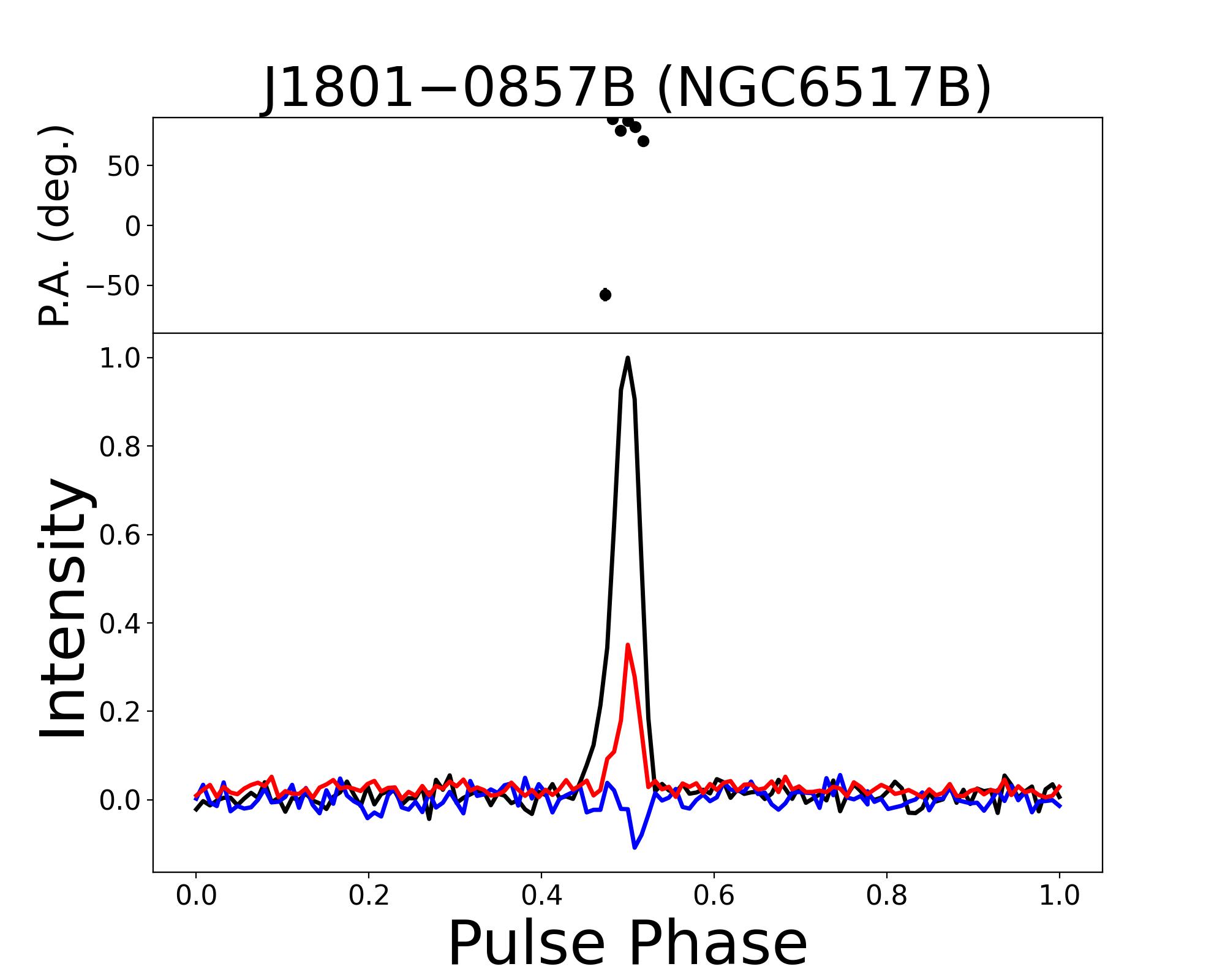}
    \end{minipage}%
    \hfill
    \begin{minipage}{0.33\textwidth}
        \includegraphics[width=\linewidth]{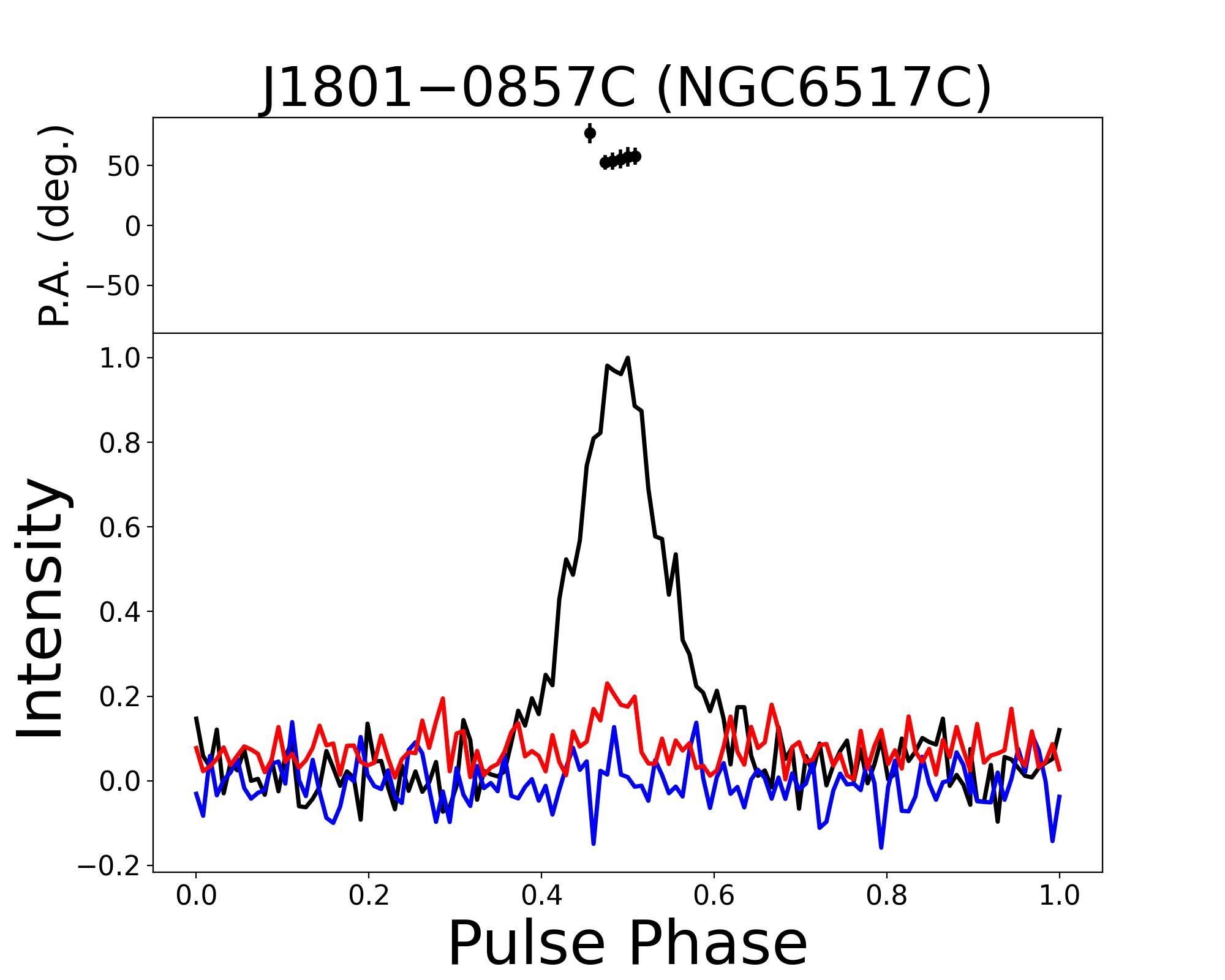}
    \end{minipage}%
    \vspace{5pt}
    \centering
    \begin{minipage}{0.33\textwidth}
        \includegraphics[width=\linewidth]{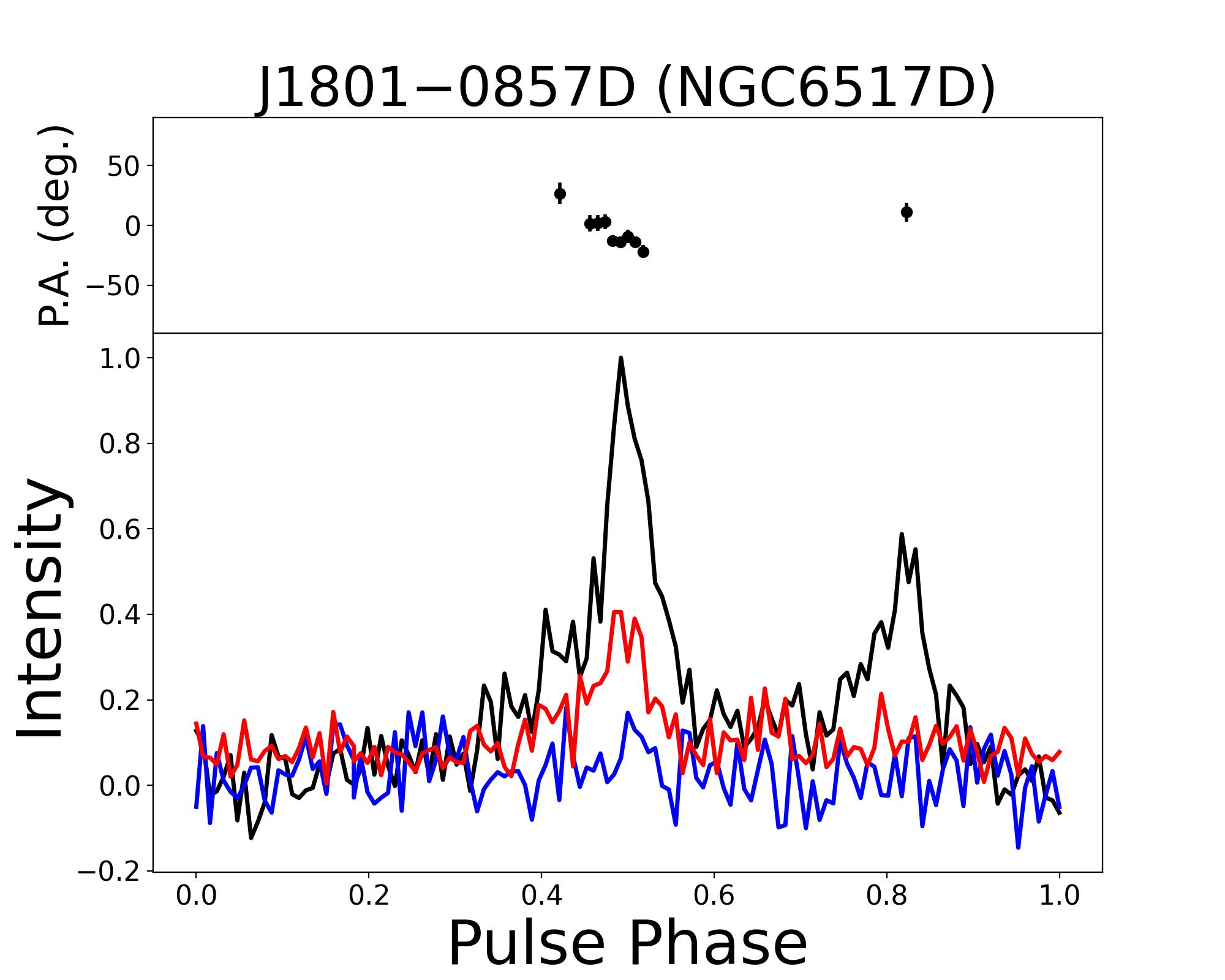}
    \end{minipage}%
    \hfill
    \begin{minipage}{0.33\textwidth}
        \includegraphics[width=\linewidth]{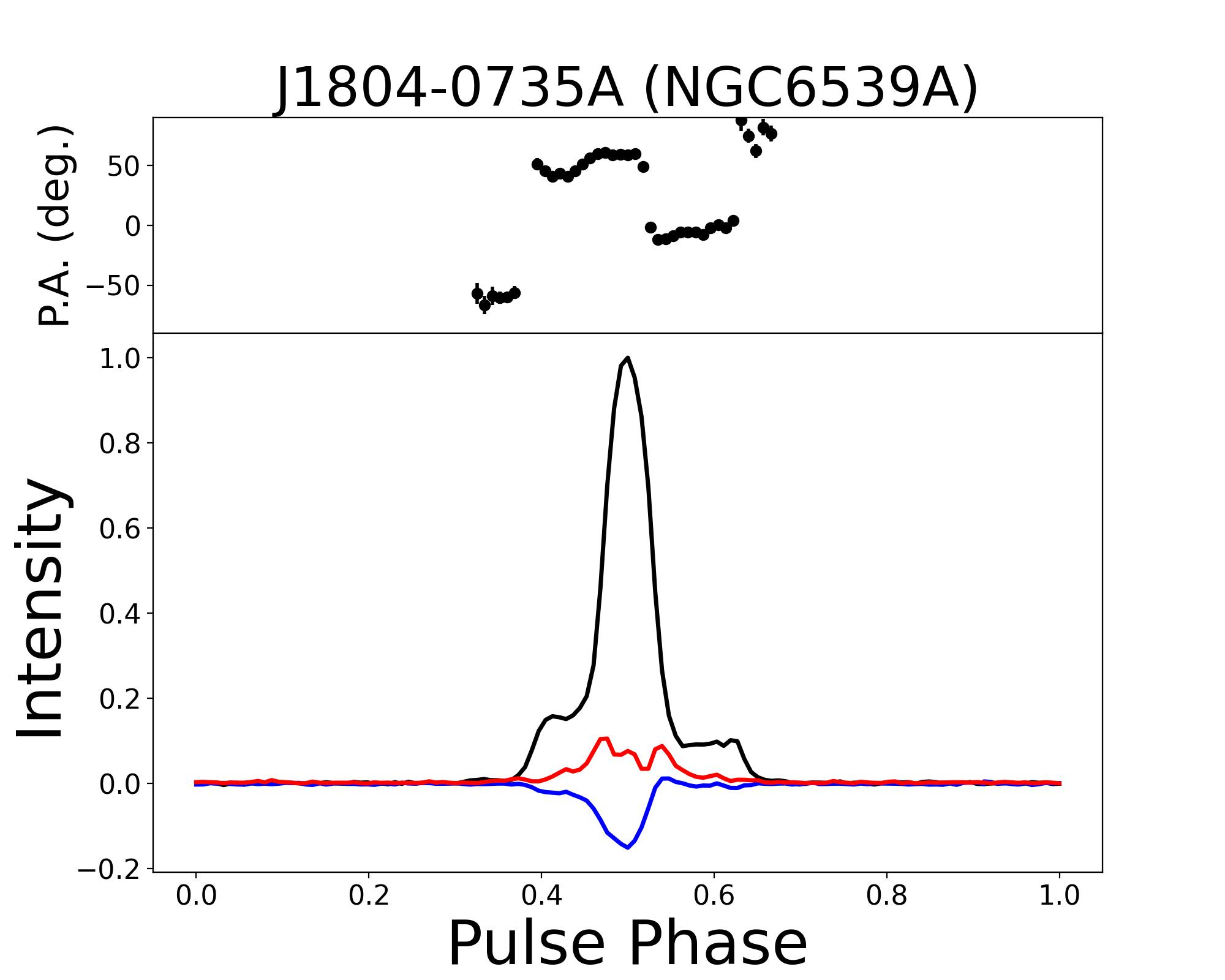}
    \end{minipage}%
    \hfill
    \begin{minipage}{0.33\textwidth}
        \includegraphics[width=\linewidth]{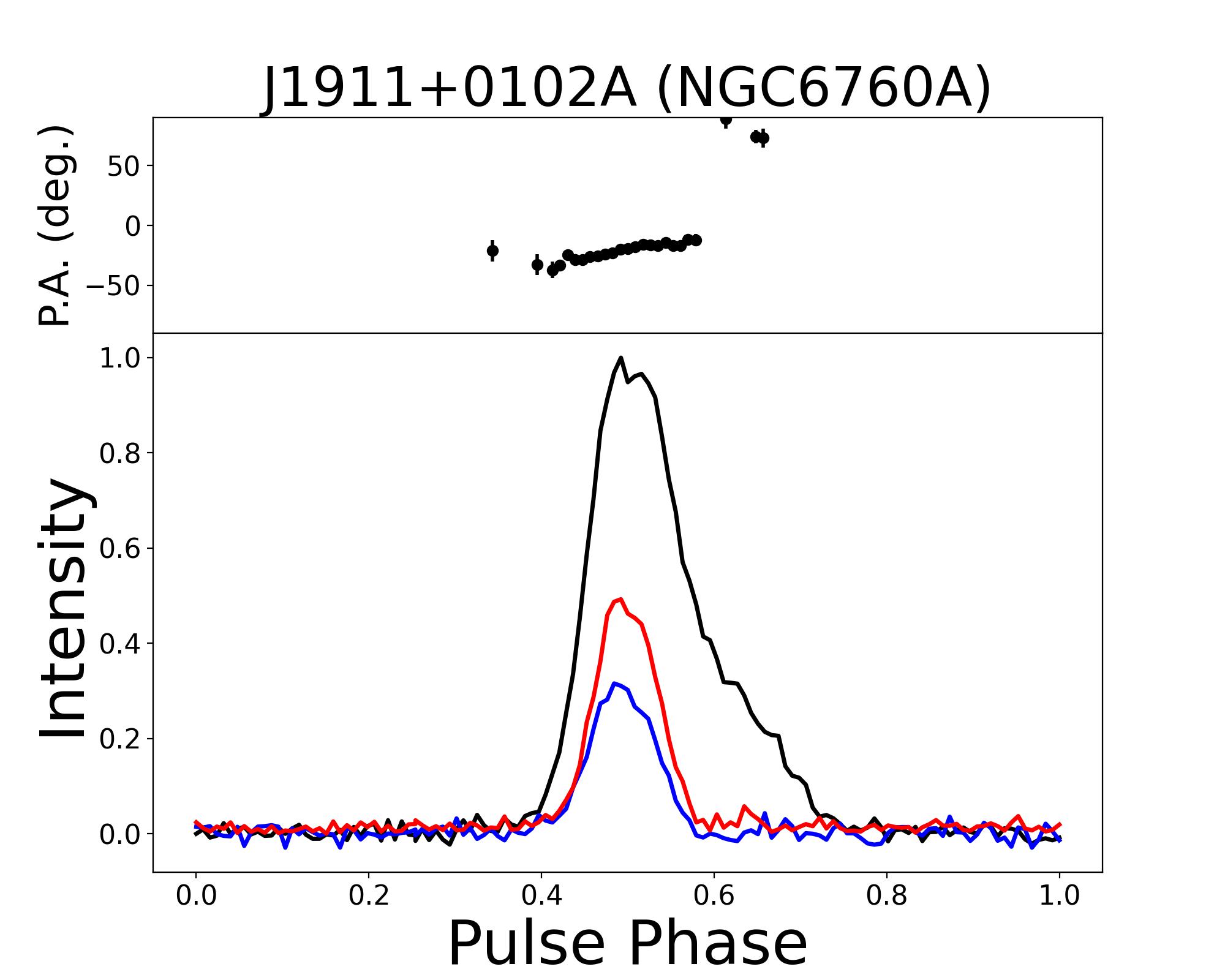}
    \end{minipage}%
    \vspace{5pt}
    \caption{Polarization profiles of 25 GC pulsars. For each profile, the upper panel represents the PA. In the lower panel, the black line represents the total intensity, the red line represents the linear polarization, and the blue line represents the circular polarization. The pulsar name are in top of each panel.}
\end{figure}

\clearpage
\begin{figure}[h!]

    \begin{minipage}{0.33\textwidth}
        \includegraphics[width=\linewidth]{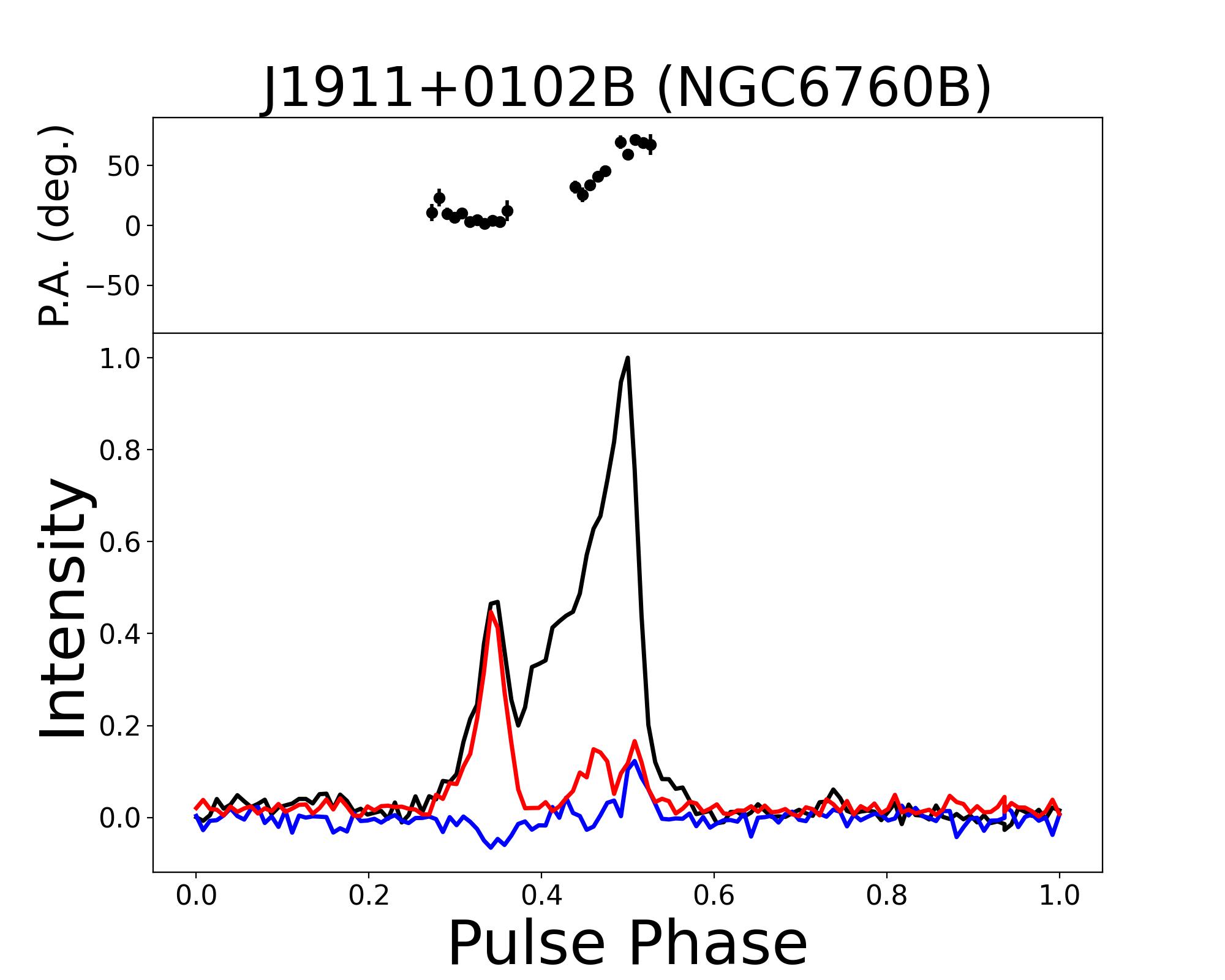}
    \end{minipage}%
    \hfill
    \begin{minipage}{0.33\textwidth}
        \includegraphics[width=\linewidth]{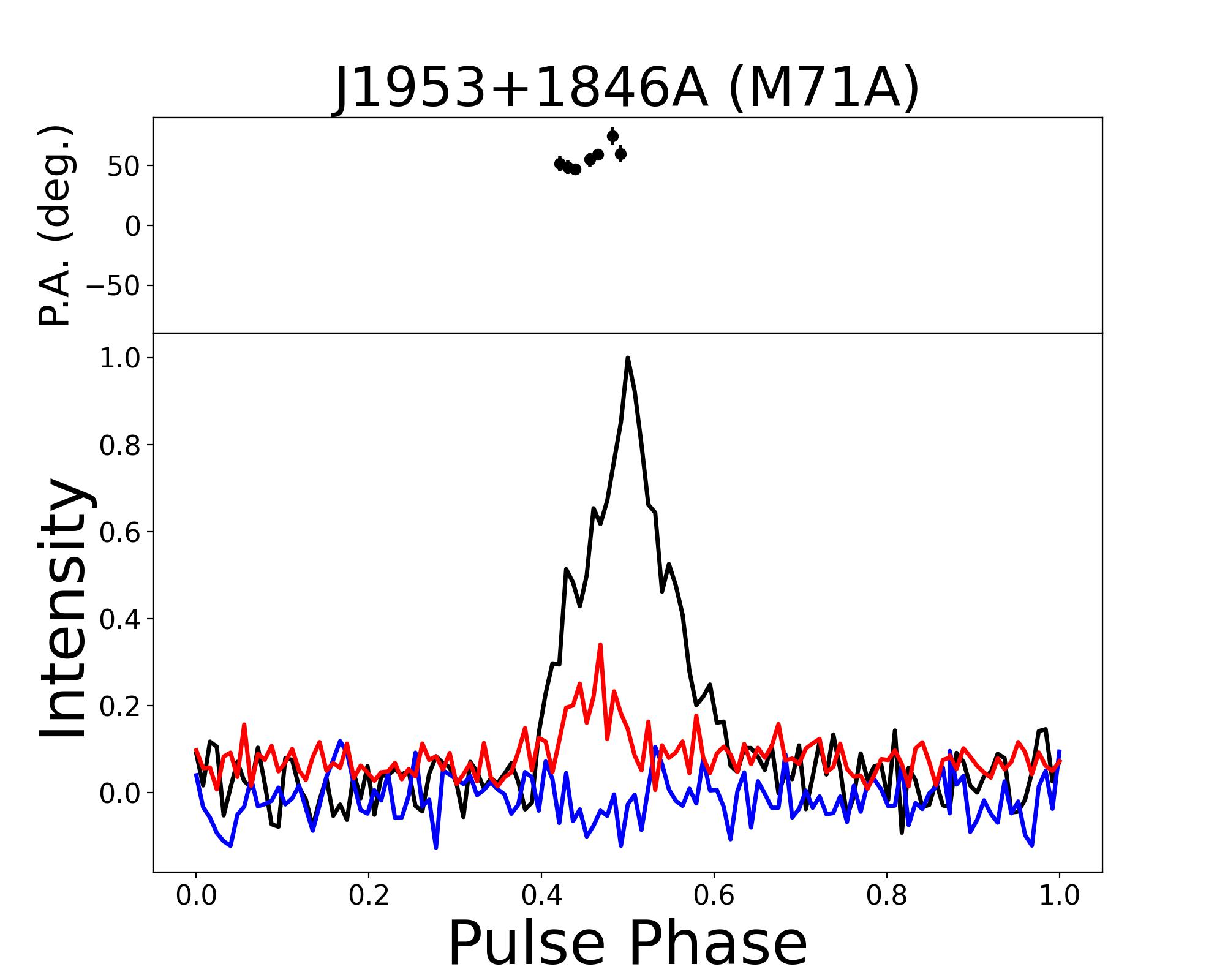}
    \end{minipage}%
    \hfill
    \begin{minipage}{0.33\textwidth}
        \includegraphics[width=\linewidth]{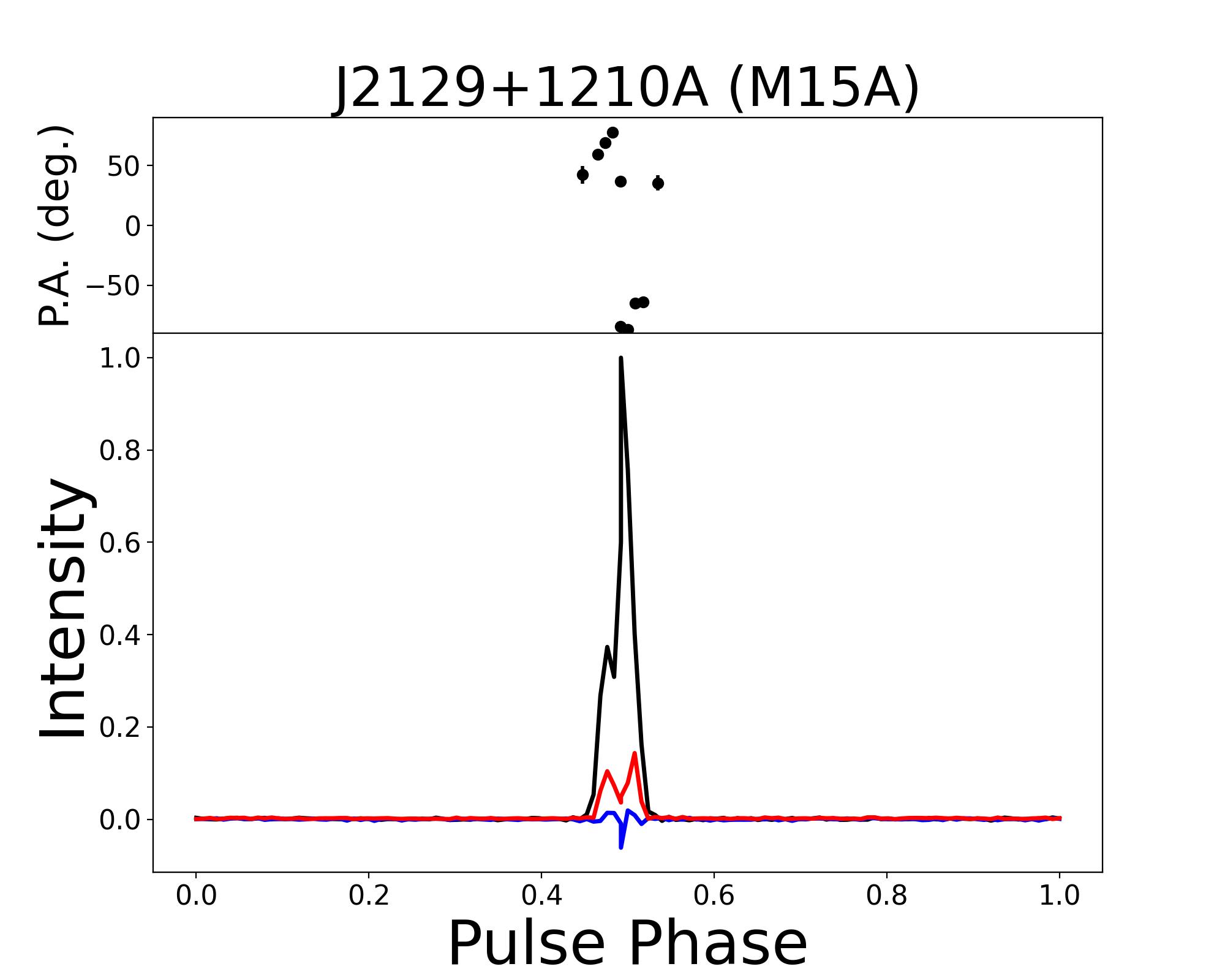}
    \end{minipage}%
    \vspace{5pt}
    \begin{minipage}{0.33\textwidth}
        \includegraphics[width=\linewidth]{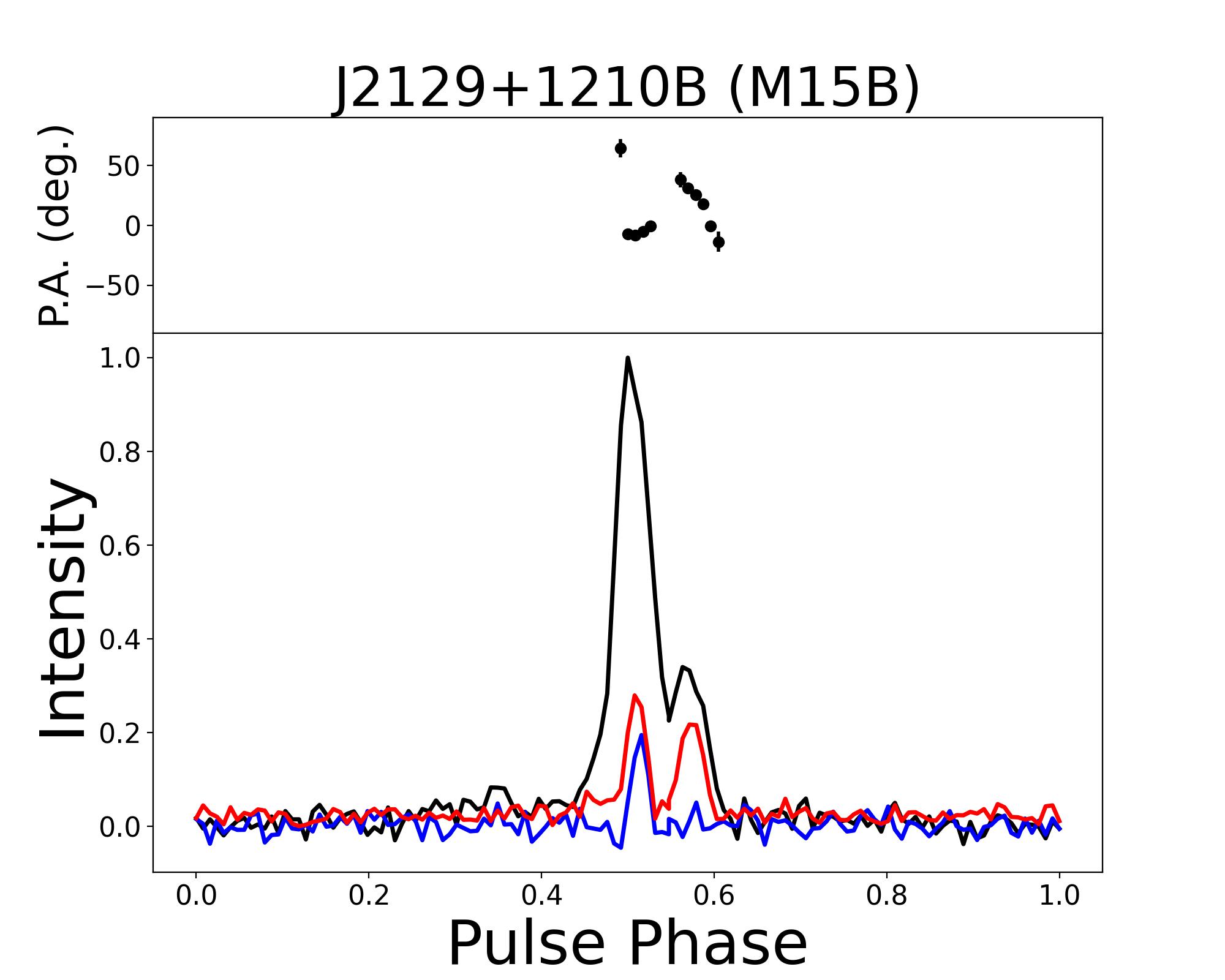}
    \end{minipage}%
    \hfill
    \begin{minipage}{0.33\textwidth}
        \includegraphics[width=\linewidth]{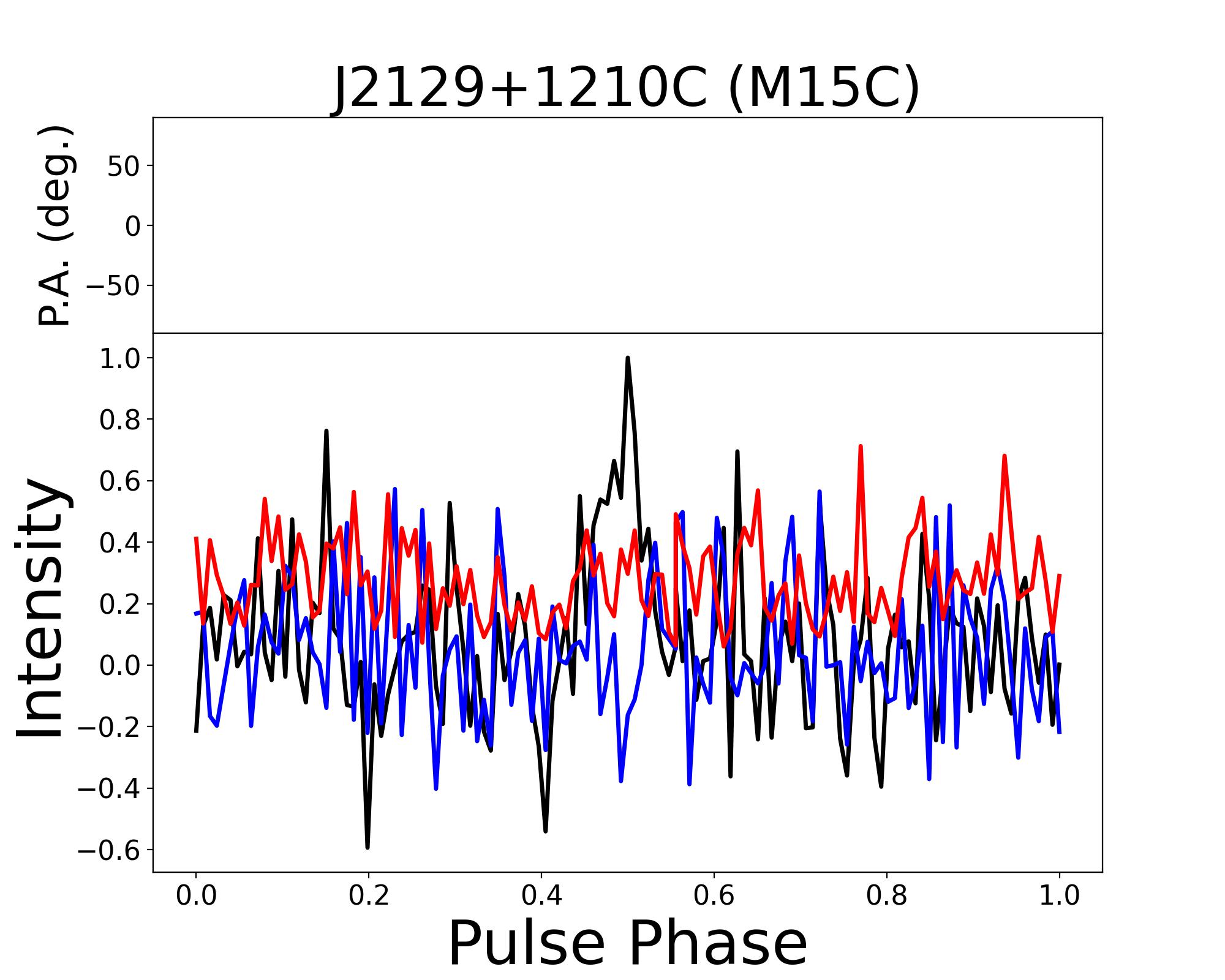}
    \end{minipage}%
    \hfill
    \begin{minipage}{0.33\textwidth}
        \includegraphics[width=\linewidth]{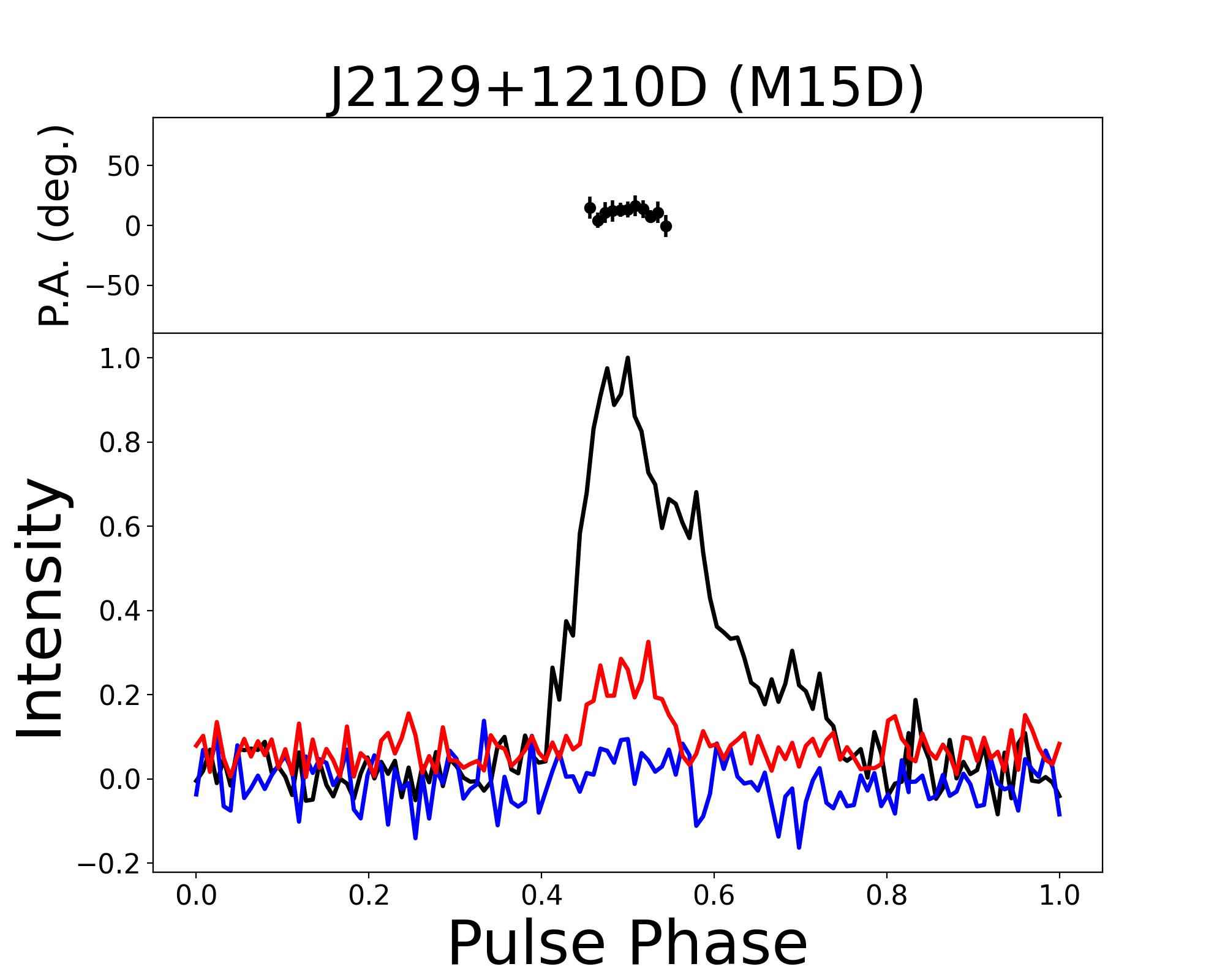}
    \end{minipage}%
    \vspace{5pt}
    \begin{minipage}{0.33\textwidth}
        \includegraphics[width=\linewidth]{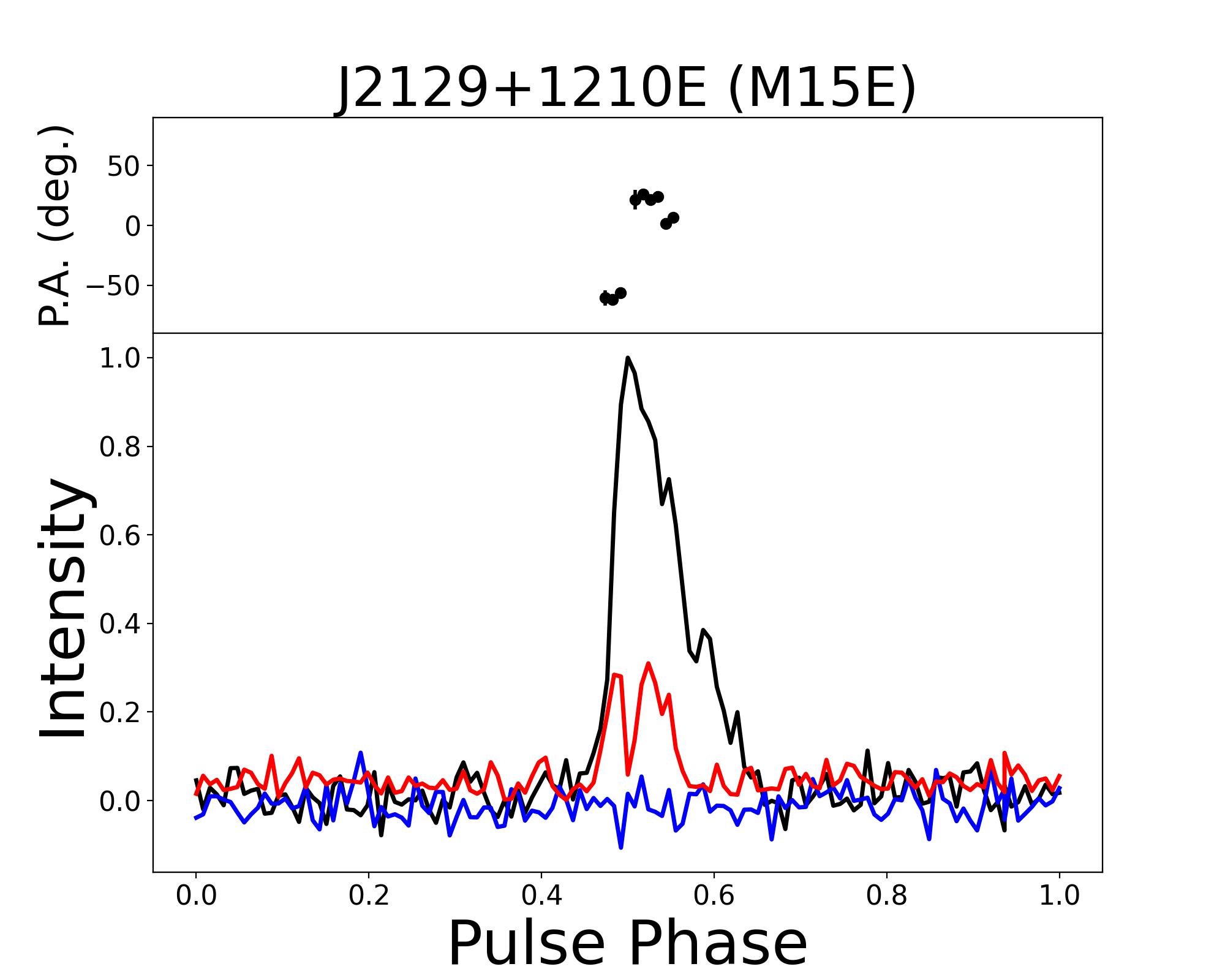}
    \end{minipage}%
    \hfill
    \begin{minipage}{0.33\textwidth}
        \includegraphics[width=\linewidth]{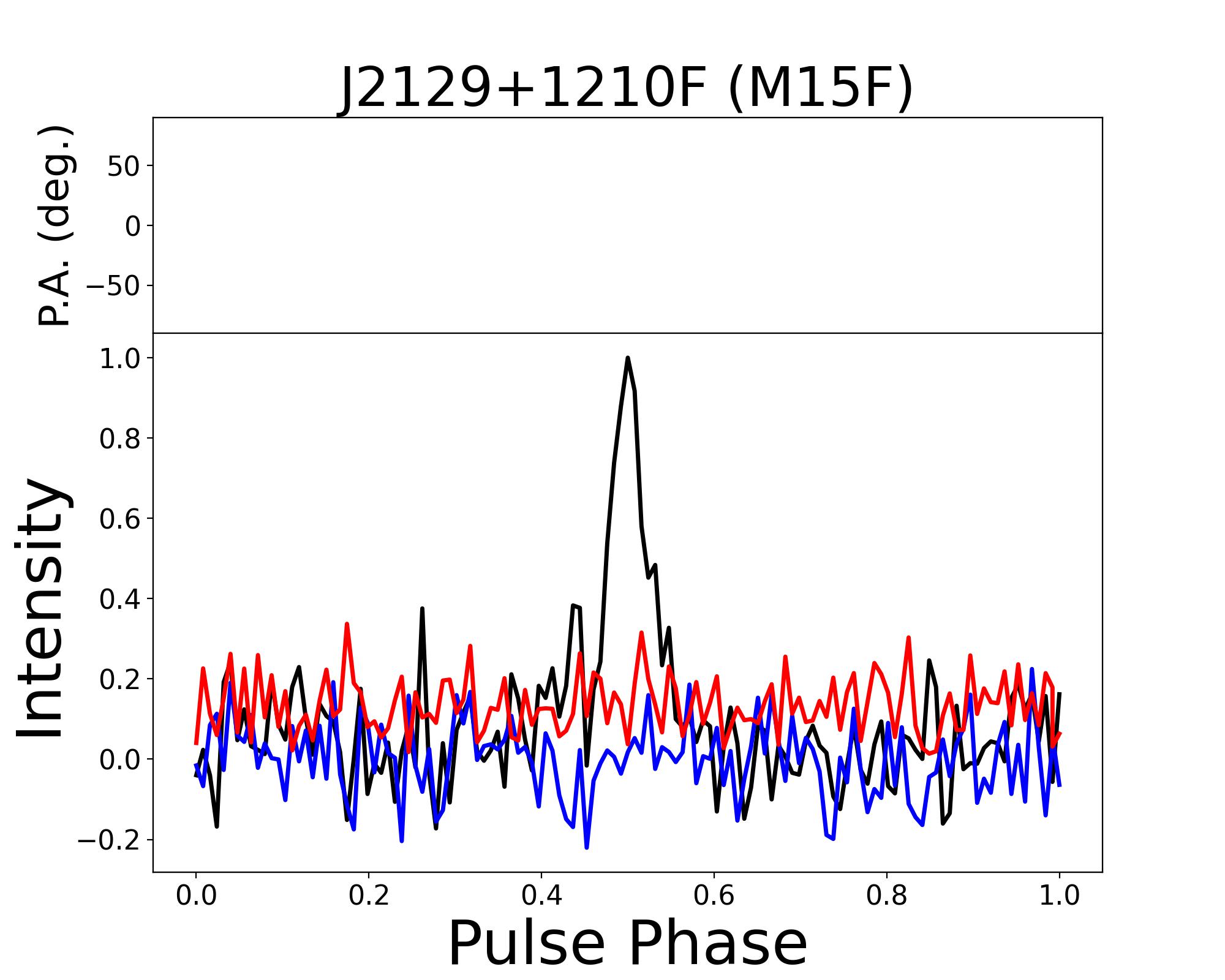}
    \end{minipage}%
    \hfill
    \begin{minipage}{0.33\textwidth}
        \includegraphics[width=\linewidth]{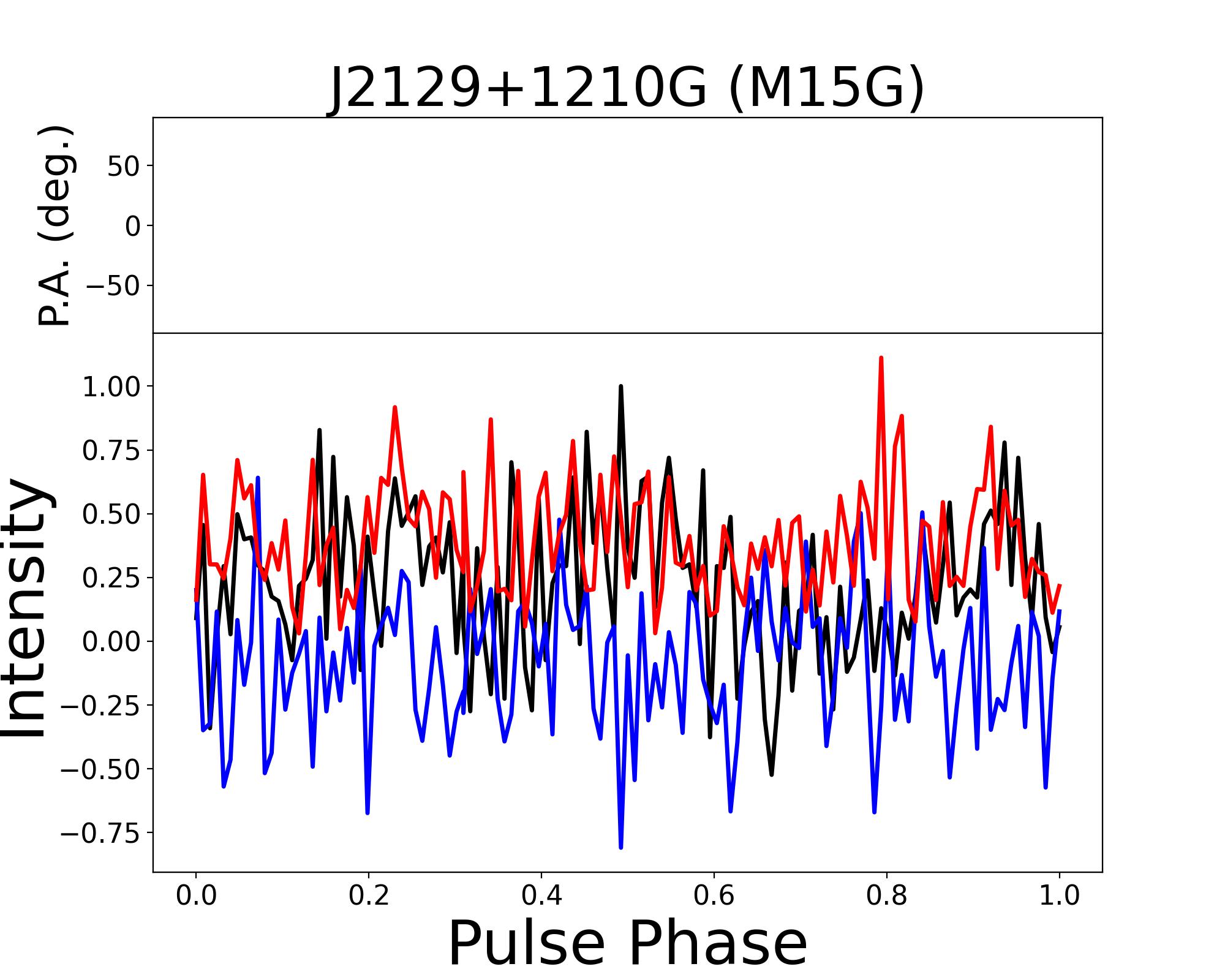}
    \end{minipage}%
    \vspace{5pt}
    \begin{center}
        \begin{minipage}{0.33\textwidth}
            \includegraphics[width=\linewidth]{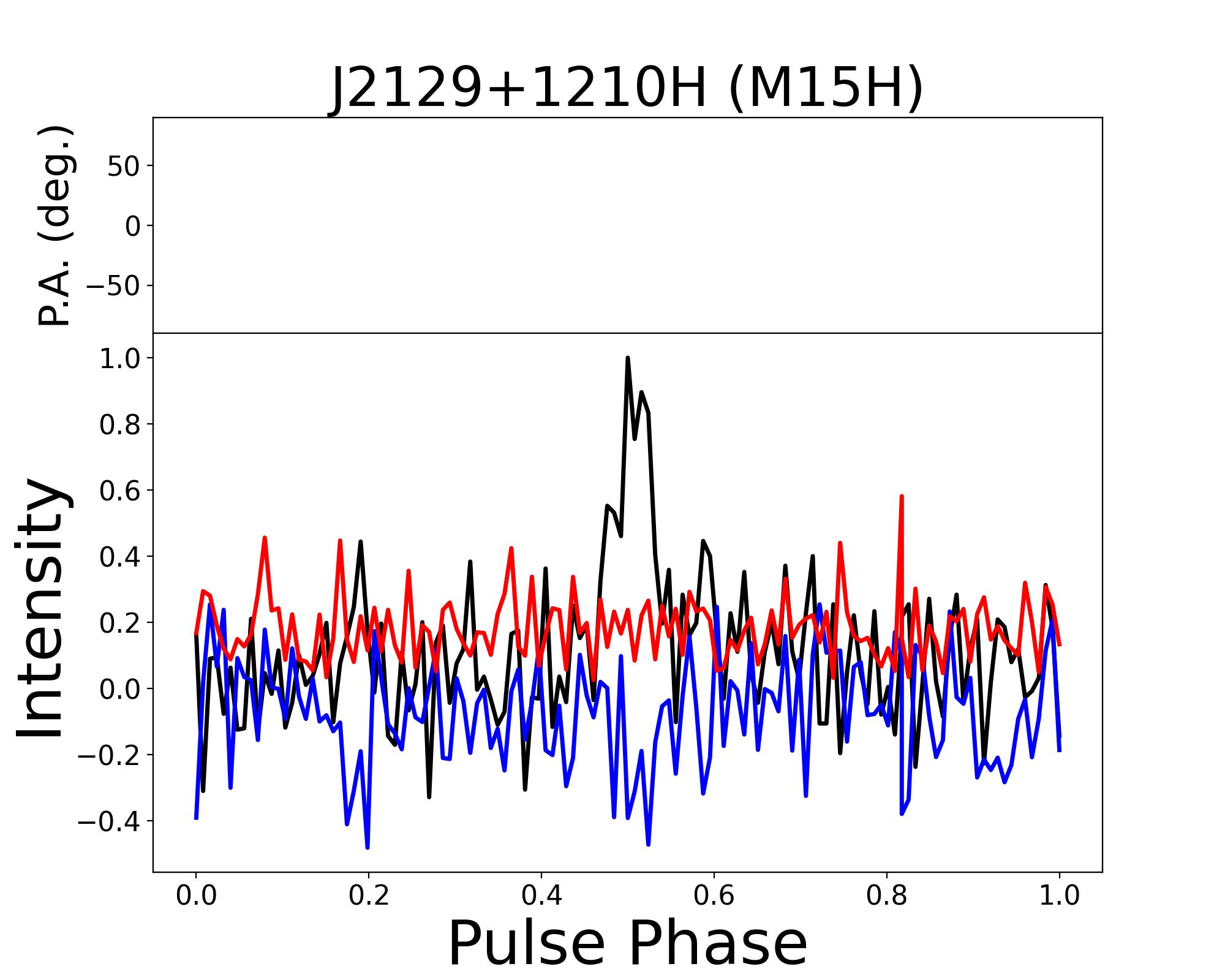}
        \end{minipage}%
    \end{center}
    \vspace{10pt}
    \renewcommand{\thefigure}{2}
    \addtocounter{figure}{-1} 
    \caption{Continued polarization profiles of 25 GC pulsars. For each profile, the upper panel represents the PA. In the lower panel, the black line represents the total intensity, the red line represents the linear polarization, and the blue line represents the circular polarization. The pulsar name are in top of each panel.}
\label{fig2}
\end{figure}


\begin{table*}
\begin{tiny}
\caption{Polarization Measurements for GC pulsars}
\label{tab2}
\begin{tabular}{lccccccccccccc}
\hline\hline
PSR &  & Date & Integration Time & P   & DM                             & PB   & S/N & $W_{10}$ & $W_{50}$ & RM                             & $L/I$   & $V/I$ & $|V|/I$\\
    &  &      &             s    & ms  & $\text{pc}\cdot\text{cm}^{-3}$ & days &     & $^\circ$ & $^\circ$ & $\text{rad}\cdot\text{m}^{-2}$ & \%      & \%    & \%     \\
\hline
J1312+1810A & M53A & 20241010 & 15660 & 33.16 & 25.03 & 255.86 & 341.0 & 40.1 & 33.1 & -2(1) & 57 & -21 & 22\\
\hline
J1342+2822A & M3A & 20240921 & 10920 & 2.54 & 26.43 & 0.14 & 182.7 & 43.8 & 25.2 & 8(1) & 29 & 1 & 3\\
J1342+2822B & M3B & 20241003 & 9900 & 2.39 & 26.15 & 1.42 & 64.3 & 157.9 & 22.0 & 16(1) & 21 & 3 & 8\\
J1342+2822D & M3D & 20240924 & 11700 & 5.44 & 26.37 & 128.75 & 46.9 & 48.6 & 20.3 & 9(2) & 47 & 0 & 8\\
\hline
J1518+0204A & M5A & 20220216 & 7200 & 5.55 & 30.05 & i & 909.5 & 50.2 & 24.0 & -3(20) & 19 & 3 & 5\\
J1518+0204B & M5B & 20220216 & 7200 & 2.95 & 29.47 & 6.86 & 124.0 & 193.3 & 50.6 & -4(6) & 12 & -5 & 7\\
J1518+0204C & M5C & 20220211 & 7200 & 2.82 & 29.31 & 0.087 & 225.7 & 188.1 & 155.6 & -1(2) & 12 & 0 & 2\\
J1518+0204D & M5D & 20220211 & 7200 & 2.90 & 29.37 & 1.22 & 39.5 & 290.3 & 73.0 & - & 19 & 8 & 16\\
J1518+0204E & M5E & 20220216 & 7200 & 3.18 & 29.31 & 1.10 & 302.0 & 260.4 & 99.3 & -2(8) & 48 & 4 & 5\\
\hline
J1801−0857A & NGC6517A & 20221231 & 8400 & 7.18 & 182.66 & i & 69.3 & 243.1 & 73.7 & 187(2) & 18 & 2 & 8\\
J1801−0857B & NGC6517B & 20221231 & 8400 & 28.96 & 182.40 & 59.84 & 82.6 & 25.4 & 12.8 & 187(1) & 26 & -5 & 7\\
J1801−0857C & NGC6517C & 20221228 & 8400 & 3.74 & 182.36 & i & 45.2 & 327.0 & 45.4 & 212(2) & 16 & 1 & 6\\
J1801−0857D & NGC6517D & 20230929 & 9000 & 4.23 & 174.54 & i & 37.8 & 2.6 & 133.8 & 189(2) & 40 & 10 & 13\\
\hline
J1804-0735A & NGC6539A & 20241009 & 3000 & 23.10 & 186.32 & 2.62 & 1320.9 & 82.5 & 21.6 & 109(1) & 13 & -12 & 13\\
\hline
J1911+0102A & NGC6760A & 20240930 & 6600 & 3.62 & 202.68 & 0.14 & 302.7 & 105.7 & 46.2 & 129(4) & 32 & 19 & 20\\
J1911+0102B & NGC6760B & 20240930 & 6600 & 5.38 & 196.69(2) & i & 152.2 & 82.7 & 24.1 & 102(1) & 29 & 1 & 8\\
\hline
J1953+1846A & M71A & 20220530 & 5285 & 4.89 & 117.39 & 0.18 & 44.8 & 245.6 & 43.9 & -480(14) & 24 & -4 & 8 \\
\hline
J2129+1210A & M15A & 20220428 & 7200 & 110.66 & 67.23 & i & 1056.3 & 23.3 & 8.8 & -72(1) & 15 & -1 & 4\\
J2129+1210B & M15B & 20220428 & 7200 & 56.13 & 67.73 & i & 96.5 & 56.0 & 17.3 & -70(2) & 27 & 5 & 9\\
J2129+1210C & M15C & 20220428 & 7200 & 30.53 & 67.10 & 0.34 & 4.1 & - & - & - & - & - & -\\
J2129+1210D & M15D & 20230428 & 7200 & 4.80 & 67.28 & i & 80.2 & 199.2 & 51.8 & -76(1) & 23 & 4 & 8\\
J2129+1210E & M15E & 20231206 & 9664 & 4.65 & 66.59 & i & 80.0 & 113.8 & 28.8 & -73(6) & 26 & -2 & 5\\
J2129+1210F & M15F & 20231206 & 9664 & 4.03 & 65.60 & i & 17.5 & - & 18.0 & - & 25 & 4 & 6\\
J2129+1210G & M15G & 20231206 & 9664 & 37.66 & 66.40 & i & 11.5 & - & - & - & - & - & - \\
J2129+1210H & M15H & 20231206 & 9664 & 6.74 & 67.12 & i & 10.6 & - & - & - & 30 & -37 & 41 \\
\hline
\end{tabular}
\footnotesize{* DM values are extracted from references listed in Table \ref{tab1}. Those without uncertainties are obtained by rounding the original values to two decimal places. RM values were obtained by fitting using \texttt{rmfit}. The absence of RM values indicates that the S/N is too low to determine the RM using this method.}
\end{tiny}
\end{table*}


\subsection{Statistical Analysis}

Pulsars M53A, M3A, M3B, M3D, and NGC6760B exhibit significant S-shaped PA distributions, 
which can be well-modeled using the RVM.
In contrast, more than 50\% of slow pulsars display S-shaped PA variations that are consistent with RVM \citep{Johnston2023}. 
A similar proportion was reported by \citet{Karastergiou2024} for MSPs in the GP, 
based on observations conducted with MeerKAT. 
This discrepancy may underscores potential differences in the PA distributions of pulsars in GCs compared to those in the GP, 
while we emphasized that these differences may caused by the small sample of GC pulsar, 
or more sensitive observations of a large sample of GC pulsars are needed.

\begin{figure}
\centering
\includegraphics[width=\textwidth, angle=0]{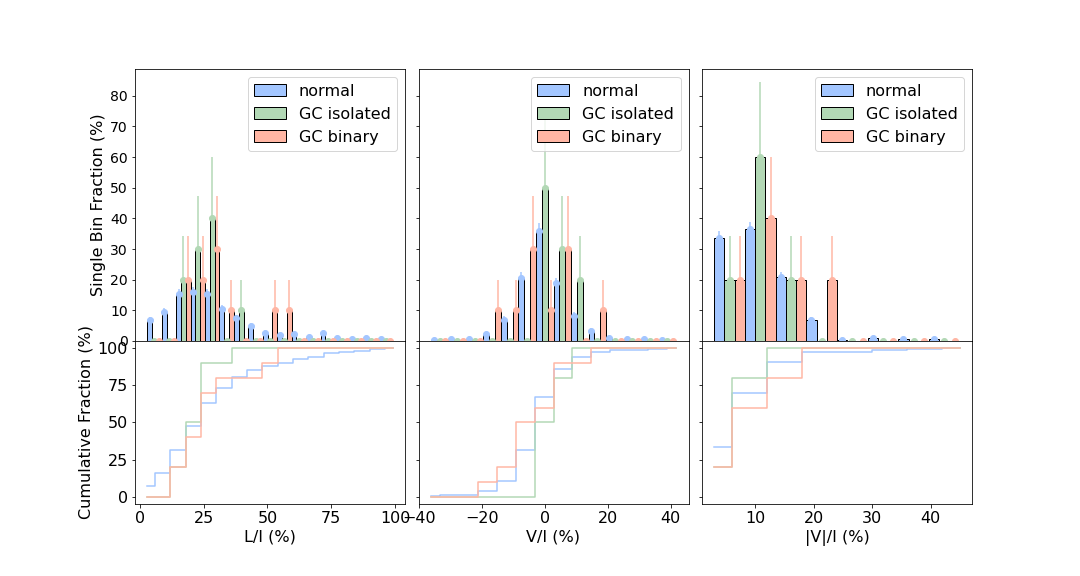}
\caption{Histograms and cumulative distribution of polarization ratios for different pulsar categories. Normal pulsars are extracted from \citealt{Wang2023}, which are pulsars with periods greater than 50 ms and located in the GP.}
\label{fig3}
\end{figure}

The average ratios of linear, circular and absolute circular polarization of the 25 GC pulsars are 26\%, 1\%, 10\%, respectively.
\citet{Oswald2023} reported average polarization fractions of 28\%, 5\%, and 32\% for linear, circular, and absolute circular polarization, respectively.
This study was based on a sample of 271 pulsars, primarily slow pulsars located in the GP and observed with Parkes around 1400\,MHz.
In comparison, the average circular and absolute circular polarization fractions measured in this work are significantly lower than those reported for slow pulsars by \citet{Oswald2023}. 
This discrepancy may reflect intrinsic differences in the polarization properties of pulsars in GCs compared to those in the GP. 
Again we emphasize that the small sample of GC pulsar may lead to these differences.
Figure \ref{fig3} depicts histogram and cumulative distribution of polarization ratios for slow pulsars, GC isolated pulsars and GC binary pulsars.
Although the evolutionary processes of GC pulsars are influenced by the internal dynamics of GCs, 
differing from the evolutionary processes of MSPs in the Galactic disk, 
the results indicate that their polarization ratio distributions are consistent.
The polarization ratio distribution of GC isolated pulsars appears slightly different from the others.
We also suggest that our limited number of samples (25 pulsars) is the main reason.
Future observations of more GC pulsars will be conducted to test this hypothesis.

\begin{figure}
\centering
\includegraphics[width=\textwidth, angle=0]{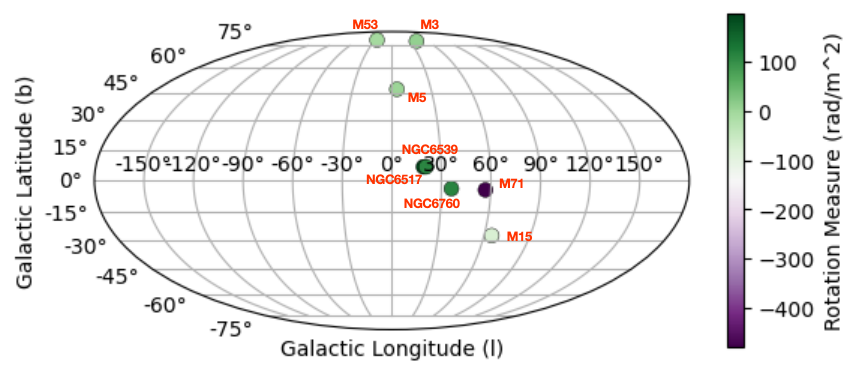}
\caption{GC distribution with RM in Galactic coordinate system. The RM value for each GC is calculated as the average RM of pulsars within.}
\label{fig4}
\end{figure}

RM values are also measured and listed in Tab \ref{tab2}.
Fig \ref{fig4} shows the average RM of pulsars in each GCs in the Galactic coordinate system.
GCs closer to the GP, namely M53 and M3, show larger RM values,
whereas those farther from the GP, such as M71 and NGC6517, show relatively lower RM values.
This result is consistent with the Galactic RM distribution reported by \citealt{Hutschenreuter2022}.
Future work will involve measurements of more GC pulsars.


\section{Conclusions}
\label{sect:conclusion}

In this study, 
we measured polarization profiles of 25 pulsars in GCs with FAST, with 15 profiles being measured for the first time.
The diversity of polarization profiles shows complex magnetic structure and emission pattern of pulsars.
Power-law index of $W_{10}$ and $W_{50}$ to the period for MSPs in GCs and GP are -0.268 and -0.330, respectively, being consistent with those of normal pulsars.
Approximately 20\% of these 25 pulsars exhibit S-shaped PA, namely M3A, M3B, M3D, M53A and NGC6760A, lower than that of normal pulsars (\citealt{Johnston2023}).
Linear, circular and absolute circular polarization ratio are measured for each pulsars.
M53A present a 57\% linear polarization ratio, being highest among these 25 GC pulsars.
M15H present a 42\% circular polarization ratio, which is the highest among these 25 pulsars.
The average ratios of circular and absolute circular polarization of these GC pulsars are -1\% and 10\% respectively, 
lower than normal pulsars measured with Parkes (\citealt{Oswald2023}), which are 5\% and 32\%, respectively.
Distribution of polarization ratios of the 25 pulsars consistent with normal pulsars derived from {\citealt{Wang2023}},
indicating that the high stellar density environment in GCs may not affect the beam evolution of MSPs.
%

The RM ranges were given by measuring the signal of pulsars in GCs.
The RM values of pulsars in each GC are similar.
The RM values or ranges for M3, M5, M15, M71, M53, NGC6517, NGC6539, NGC6760 are 
8(1) to 16(1) rad $m^{-2}$, 
-1(2) to -4(6) rad $m^{-2}$, 
-70(2) to -76(1) rad $m^{-2}$,
-480(14) rad $m^{-2}$ (M71A only),
-2(1) rad $m^{-2}$ (M53A only),
187(1) to 212(2) rad $m^{-2}$, 
109(1) rad $m^{-2}$ (NGC6539A only), 
and 102(1) to 129(4) rad $m^{-2}$.
%
The GCs closer to the GP tend to have larger RM.
This is consistent with previous study (\citealt{Hutschenreuter2022}). 

We presented the polarization profiles of 25 pulsars, 26\% of all the 97 GC pulsars in the FAST sky.
%
Besides our study, on-going works are also on M2 (Li et al. in prep), M13 (Wang et al. in prep), 
M14 (Liu et al. in prep), and NGC6749 (Freire et al. in prep).
%
%
%
More GC pulsars still need public timing solutions (e.g., M92B\footnote{\url{https://fast.bao.ac.cn/cms/article/82}}), or, with too high DM to be observed by FAST at L-band (e.g., some pulsars in Glimpse C01\footnote{\url{https://www.trapum.org}}).
Polarization studies on some of these pulsars will be in our next paper.

\begin{acknowledgements}
This work is supported by National Key R\&D Program of China, No. 2022YFC2205202 and the National Natural Science Foundation of China (NSFC) under grant Nos. 12003047 and 12173053. Zhichen Pan is supported by the Youth Innovation Promotion Association of the Chinese Academy of Sciences (CAS) (id. Y2022027), the CAS “Light of West China” Program, and the Chinese Academy of Sciences President’s International Fellowship Initiative (grant No. 2021FSM0004).
This work made use of data from the Five-hundred-meter Aperture Spherical radio Telescope (FAST), a Chinese national mega-science facility operated by the National Astronomical Observatories, Chinese Academy of Sciences. We appreciate the availability of publicly released FAST data and are grateful to Dejiang Yin, Yuxiao Wu, Baoda Li, and Yujie Lian for providing related data.
\end{acknowledgements}



\clearpage
\bibliographystyle{raa}
\bibliography{bibtex}

\label{lastpage}

\end{document}